\documentclass[11pt, a4paper]{article}
\usepackage{graphicx}
\usepackage{amssymb}
\usepackage{amsmath}
\usepackage{bm}
\usepackage{color}
\usepackage{theorem}
\usepackage{subcaption}
\usepackage{braket}
\usepackage{multirow}
\usepackage{caption}
\usepackage{arydshln}

\usepackage[sort&compress,numbers, merge]{natbib}

\definecolor{Orange}{cmyk}{0,0.61,0.87,0}
\definecolor{JungleGreen}{cmyk}{0.99,0,0.52,0}
\definecolor{OliveGreen}{cmyk}{0.64,0,0.95,0.40}
\definecolor{Brown}{cmyk}{0,0.81,1,0.60}
\definecolor{RoyalBlue}{cmyk}{0.71,0.53,0,0.12}

\renewcommand{\arraystretch}{1.3}

\usepackage[colorlinks=true, linkcolor=OliveGreen, citecolor=RoyalBlue, urlcolor=Brown]{hyperref}

\begin{document}

\begin{titlepage}

\begin{flushright}
{\tt 
UT-19-17
}
\end{flushright}

\vskip 1.35cm
\begin{center}

{\Large
{\bf
Predictions for the neutrino parameters in the minimal model extended by linear combination of U(1)$_{L_e-L_\mu}$, U(1)$_{L_\mu-L_\tau}$ and U(1)$_{B-L}$ gauge symmetries
}
}

\vskip 1.2cm

Kento Asai

\vskip 0.4cm

{\it Department of Physics, University of Tokyo, Bunkyo-ku, Tokyo
 133--0033, Japan} 

\date{\today}

\vskip 1.5cm

\begin{abstract}
We study the minimal extensions of the Standard Model by a linear combination of U(1)$_{L_e-L_\mu}$, U(1)$_{L_\mu-L_\tau}$ and U(1)$_{B-L}$ gauge symmetries, where three right-handed neutrinos and one U(1)-breaking SU(2)$_L$ singlet or doublet scalar are introduced. 
Because of the dependence on the lepton flavor, the structures of both Dirac and Majorana mass matrices of neutrinos are restricted. 
In particular, the two-zero minor and texture structures in the mass matrix for the active neutrinos are interesting. 
Analyzing these structures, we obtain uniquely all the neutrino parameters, namely the Dirac CP phase $\delta$, the Majorana CP phases $\alpha_{2,3}$ and the mass eigenvalues of the light neutrinos $m_i$ as functions of the neutrino mixing angles $\theta_{12}$, $\theta_{23}$, and $\theta_{13}$, and the squared mass differences $\Delta m^2_{21}$ and $\Delta m^2_{31}$.
In 7 minimal models which are consistent with the recent neutrino oscillation data, we also obtain the predictions for the sum of the neutrino masses $\Sigma_i m_i$ and the effective Majorana neutrino mass $\braket{m_{\beta \beta}}$ and compare them with the current experimental limits. 
In addition, we also discuss the implication of our results for leptogenesis.

\end{abstract}

\end{center}
\end{titlepage}

\section{Introduction}
\label{sec:intro}

Two decades have passed since the Super-Kamiokande observed the neutrino oscillation in 1998. 
From that time on, a lot of experiments, such as T2K~\cite{Abe:2017vif,Abe:2018wpn}, KamLAND~\cite{KamLAND-Zen:2016pfg}, and so on, are conducted, and then the neutrino mixing angles and the squared mass differences are vigorously measured. 
The observations of neutrino oscillations claim that neutrinos have non-zero masses and we have to modify the neutrino sector in the Standard Model (SM). 
During these two decades, a lot of modifications were proposed to solve this puzzle. 
The simplest way to explain the neutrino masses is to introduce right-handed neutrinos into the SM. 
If they couple to the SM leptons and the Higgs field, the active neutrinos acquire the Dirac masses as the other fermions do. 
Moreover, the smallness of the neutrino masses can be naturally understood by the seesaw mechanism~\cite{Minkowski:1977sc,Yanagida:1979as,GellMann:1980vs,Mohapatra:1979ia}.
For instance, if neutrinos are Majorana particles, and the Majorana masses are much heavier than the Dirac masses, then the tiny neutrino masses are realized by the type-I seesaw mechanism.

On the other hand, it is often considered to extend the SM gauge sector by U(1) gauge symmetries~\cite{Foot:1990mn,He:1991qd,Foot:1994vd,He:1990pn}. 
For instance, it is well known that U(1)$_{B-L}$ can be introduced to the SM without anomalies when three right-handed neutrinos are introduced to the SM. 
In addition, it is well known that the U(1)$_{L_\alpha-L_\beta}$ gauge symmetries, where $L_\alpha$ represents the lepton number of generation associated with $\alpha\ (=e, \mu, \tau)$, can be also introduced to the SM without regard to whether three right-handed neutrinos are introduced or not.
Above all, the U(1)$_{L_\mu-L_\tau}$ gauge symmetry is often discussed in the contexts of the muon $g-2$ anomaly~\cite{Bennett:2006fi,Jegerlehner:2009ry,Davier:2010nc,Hagiwara:2011af,Baek:2001kca,Ma:2001md}, flavor physics~\cite{Altmannshofer:2014cfa,Crivellin:2015mga}, dark matter~\cite{Kim:2015fpa,Baek:2015fea,Patra:2016shz,Biswas:2016yan,Biswas:2016yjr,Biswas:2017ait,Foldenauer:2018zrz}, and so on.
Other recent related studies on the gauged U(1)$_{L_\alpha-L_\beta}$  models are found in Refs.~\cite{Heeck:2011wj,Harigaya:2013twa,delAguila:2014soa,Fuyuto:2014cya,Araki:2015mya,Elahi:2015vzh,Fuyuto:2015gmk,Altmannshofer:2016oaq,Ibe:2016dir,Kaneta:2016uyt,Araki:2017wyg,Hou:2017ozb,Chen:2017cic,Elahi:2017ppe,Chen:2017gvf,Baek:2017sew,Gninenko:2018tlp,Wise:2018rnb,Nomura:2019uyz,Nomura:2018vfz,Bauer:2018onh,Arcadi:2018tly,Kamada:2018zxi,Liu:2018xsw,Bauer:2018egk,Nomura:2018cle,Banerjee:2018mnw,Chun:2018ibr,Escudero:2019gzq}.

In models extended by a lepton flavor-dependent U(1) gauge symmetry (hereafter written as U(1)$_{Y'}$), the structure of the neutrino mass matrix is tightly restricted. 
Then if we extend only the gauge sector and the extra U(1)$_{Y'}$ gauge symmetry is preserved, the neutrino mass matrix is too sparse to explain the neutrino oscillation data, and therefore we need to break the U(1)$_{Y'}$. 
To do so, one usually introduces a SU(2)$_L$ singlet or doublet scalar with a non-zero U(1)$_{Y'}$ charge such that its VEV spontaneously breaks the U(1)$_{Y'}$ gauge symmetry and gives a mass to the U(1)$_{Y'}$ gauge boson. 
We, however, note that even though the U(1)$_{Y'}$ gauge symmetry is spontaneously broken, the structure of the neutrino mass matrix is still highly constrained if we introduce only one U(1)$_{Y'}$-breaking scalar field and consider only renormalizable interactions. 
Particularly, in some U(1)$_{Y'}$ cases, including U(1)$_{L_\mu-L_\tau}$, (the inverse of) the neutrino mass matrix has zeros in its two components, and such a structure is called two-zero texture (minor).
For previous studies on the neutrino sector of the gauged U(1)$_{L_\alpha-L_\beta}$ models, see Refs~\cite{Branco:1988ex,Choubey:2004hn,Araki:2012ip,Heeck:2014sna,Baek:2015mna,Crivellin:2015lwa,Plestid:2016esp,Lee:2017ekw,Asai:2017ryy,Asai:2018ocx,Dev:2017fdz}.
In Refs.~\cite{Araki:2012ip,Heeck:2014sna}, the relations between lepton flavor-dependent U(1) gauge symmetries and structures of the neutrino mass matrix are comprehensively discussed and, previously, in Refs.~\cite{Lavoura:2004tu,Ma:2005py,Lashin:2007dm}, the relations between two-zero minor and texture structures and constrains on the neutrino parameters are discussed. In Refs.~\cite{Asai:2017ryy,Asai:2018ocx}, we have discussed them in detail, and have given the predictions for unknown parameters, such as the CP phases, the neutrino masses, and the effective neutrino mass, in the case of the minimal extended model by a U(1)$_{L_\alpha-L_\beta}$ gauge symmetry, that we call ``the minimal gauged U(1)$_{L_\alpha-L_\beta}$ model'', and we have found that these models are already ruled out or are driven into a corner.

In this paper, we have extended the U(1)$_{L_\alpha-L_\beta}$ gauge symmetries to a linear combination of the U(1)$_{L_e-L_\mu}$, U(1)$_{L_\mu-L_\tau}$, and U(1)$_{B-L}$, and have found viable minimal models. 
Such a linear combination is the most general lepton flavor-dependent U(1), which can be introduced to the SM gauge sector without anomalies and is consistent with renormalizable Yukawa terms when three right-handed neutrinos are added to the SM.
This kind of U(1)s are also discussed, for instance, in Refs.~\cite{Ma:1997nq,Ma:1998dp,Lee:2010hf,Nomura:2018rvy}.
We analyze the minimal models with two-zero minor or texture structure of the neutrino mass matrix in the same way as we did in Refs.~\cite{Asai:2017ryy,Asai:2018ocx}, and then we find that only three minimal models are consistent with the recent experiments.
We obtain all the CP phases in the PMNS matrix and the mass eigenstates of the light neutrinos as functions of the neutrino oscillation parameters. 
We also discuss the implication of our results for leptogenesis and give the predictions of the baryon asymmetry of the Universe.

This paper is organized as follows.
In Sec.~\ref{sec:model}, we review the anomaly cancellation conditions and then introduce the minimal gauged U(1) models and explain only fifteen U(1) gauge symmetries realize two-zero minor (TZM) or texture (TZT) structure.
In Sec.~\ref{sec:vanishing}, we show the analyses of the TZM structure following Ref.~\cite{Asai:2017ryy,Asai:2018ocx}.
We then show in Sec.~\ref{sec:result} the predictions for the Dirac CP phase $\delta$, the sum of the neutrino masses $\Sigma_i m_i$, and the effective mass for the neutrinoless double beta decay $\braket{m_{\beta \beta}}$.
In Sec.~\ref{sec:LG}, we discuss the implications for the leptogenesis. 
Finally, our conclusions are summarized in Sec.~\ref{sec:conclusion}.
In Appendix, we show the explicit expressions for the values and equations which appear in the analyses of the neutrino mass matrices.

\section{Minimal gauged U(1) models and neutrino mass matrices} 
\label{sec:model}

\subsection{Anomaly cancellation conditions}
\label{subsec:anomaly}

For a start, we discuss the conditions that lepton flavor-dependent U(1)$_{Y'}$ gauge symmetries do not cause anomalies, where $Y'$ represents the charge of the U(1)$_{Y'}$. 
We assume as follows:
\begin{itemize}
\item All quarks have the same U(1)$_{Y'}$ charges.
\item Left-handed and right-handed fermions have the same U(1)$_{Y'}$ charges.
\item Charged leptons and neutrinos with the same flavor have the same U(1)$_{Y'}$ charge.
\item U(1)$_{Y'}$ charge of the Higgs field is zero.
\end{itemize}
Tab. \ref{tab:U1charge-x} summarizes the above assumptions.

\begin{table}[t]
\centering
\caption{U(1)$_{Y'}$ charges of SM fermions and right-handed neutrinos } 
\label{tab:U1charge-x}
\vspace{5pt}
\begin{tabular}{c|ccccc} \hline \hline
   \multirow{2}{*}{\textrm{field}} & \multirow{2}{*}{\textrm{quarks}} & \multicolumn{3}{c}{\textrm{leptons}} & \multirow{2}{*}{\textrm{Higgs}} \\ \cline{3-5}
    &  & $e,\nu_e,N_e$ & $\mu,\nu_\mu,N_\mu$ & $\tau,\nu_\tau,N_\tau$ &  \\ \hline
    \textrm{charge} & $x_q$ & $x_e$ & $x_\mu$ & $x_\tau$ & 0 \\ \hline \hline
\end{tabular}
\end{table}

Under these assumptions, the cubic, gravitational, U(1)$_{Y}\times$U(1)$_{Y'}\times$U(1)$_{Y'}$ and SU(3)$_{C}$ $\times$SU(3)$_{C}\times$U(1)$_{Y'}$ anomalies vanish.
This is because left-handed and right-handed fermions have the same U(1)$_{Y'}$ charges and the opposite contributions to these anomalies. 
Therefore, as shown in Refs.~\cite{Araki:2012ip,Heeck:2014sna,Kownacki:2016pmx}, the SU(2)$_{L}\times$SU(2)$_L\times$U(1)$_{Y'}$ and U(1)$_{Y}\times$U(1)$_{Y}\times$U(1)$_{Y'}$ anomalies lead the same nontrivial condition of the anomaly cancellation as follows:
\begin{align}
\label{eq:anomaly}
9 x_q + x_e + x_\mu + x_\tau = 0.
\end{align}
Thus, the charge $Y'$ can be written as follows:
\begin{align}
   Y' = \left\{ \begin{array}{ll} x_e L_e + x_\mu L_\mu - (x_e + x_\mu) L_\tau & \ \ \ (x_q = 0) \\ B + x_e L_e + x_\mu L_\mu - (3 + x_e + x_\mu) L_\tau & \ \ \ (x_q = \frac{1}{3}) \end{array} \right.~.
\end{align}
Therefore, only two types of lepton flavor-dependent U(1) gauge symmetries, i.e. U(1)$_{x_e L_e + x_\mu L_\mu -(x_e+x_\mu) L_\tau}$ and U(1)$_{B + x_e L_e + x_\mu L_\mu - (3+x_e+x_\mu) L_\tau}$ (hereinafter abbreviated to U(1)$_{e\mu\tau}$ and U(1)$_{Be\mu\tau}$, respectively), can be introduced to the models without anomalies.

Introducing these gauge symmetries, the mass terms of the neutrino Dirac, Majorana and charged lepton are strictly restricted. 
In the $(e,\mu,\tau)$ basis, the U(1)$_{Y'}$ charges of the neutrino Dirac Yukawa and the charged lepton terms are
\begin{align}
Q_{Y'}(\text{Dirac \& charged lepton}):~ \left\{ \begin{array}{ll}
   \begin{pmatrix}
   0 & -x_e+x_\mu & -2x_e-x_\mu \\
   x_e-x_\mu & 0 & -x_e-2x_\mu \\
   2x_e+x_\mu & x_e+2x_\mu & 0
   \end{pmatrix} & \text{for}~Y'=e\mu\tau \\
   \begin{pmatrix}
   0 & -x_e+x_\mu & -3-2x_e-x_\mu \\
   x_e-x_\mu & 0 & -3-x_e-2x_\mu \\
   3+2x_e+x_\mu & 3+x_e+2x_\mu & 0
   \end{pmatrix} & \text{for}~Y'=Be\mu\tau
   \end{array}, \right.
\label{eq:charged}
\end{align}
where the $(\alpha,\beta)$ entry in the above matrices represents the U(1)$_{Y'}$ charge of the fermion bilinear term $N_\alpha^cL_\beta$ or $e_\alpha^cL_\beta$, with $\alpha,\beta$ the flavor indices. 
On the other hand, those of the Majorana mass term of the right-handed neutrinos are
\begin{align}
Q_{Y'}(\text{Majorana}):~ \left\{ \begin{array}{ll}
   \begin{pmatrix}
   -2x_e & -x_e-x_\mu & x_\mu \\
   -x_e-x_\mu & -2x_\mu & x_e \\
   x_\mu & x_e & 2(x_e+x_\mu)
   \end{pmatrix} & \text{for}~Y'=e\mu\tau \\
   \begin{pmatrix}
   -2x_e & -x_e-x_\mu & 3+x_\mu \\
   -x_e-x_\mu & -2x_\mu & 3+x_e \\
   3+x_\mu & 3+x_e & 2(3+x_e+x_\mu)
   \end{pmatrix} & \text{for}~Y'=Be\mu\tau
   \end{array}, \right.
\label{eq:chargem}
\end{align}
where the $(\alpha,\beta)$ entry in the above matrices represents the U(1)$_{Y'}$ charge of the fermion bilinear term $N_\alpha^cN_\beta^c$. 
From Eqs.~\eqref{eq:charged} and \eqref{eq:chargem}, we find that the Dirac, charged lepton, and Majorana mass matrices are sparse, at least block diagonal, even in any $Y'$ charge assignments. Then required values of the neutrino mixing angles cannot be obtained from the simple model, as long as the extra U(1)$_{Y'}$ is preserved and only renormalizable interactions are considered. 
Therefore, we introduce one SU(2)$_L$ scalar singlet or doublet with non-zero U(1)$_{Y'}$ charge
\footnote{We can choose a SU(2)$_L$ triplet as the U(1)$_{Y'}$ breaking scalar. 
However, we find, in that case, the resultant neutrino mass matrix cannot have the TZM or the TZT structure, and then we do not consider the triplet case in this paper.}.
Its VEV leads the U(1)$_{Y'}$ symmetry breaking, and then the Majorana or Dirac mass terms are generated.
These mass matrices have various structures depending on U(1)$_{Y'}$ gauge symmetries and additional scalars, and the TZM and TZT structures are especially interesting.
This is because the TZM or TZT structure enables us to predict the low-energy neutrino parameters.
When (the inverse of) the neutrino mass matrix has more than two-zeros, the vanishing conditions lead too many constraints and there is no solution.
On the other hand, when (the inverse of) the neutrino mass matrix has one or no zero, the model has less predictive power and this case is less interesting.
That is why we focus on the TZM and TZT cases in this paper.
In Tab.~\ref{tab:tzmlist}, we summarize the U(1)$_{Y'}$s which realize the TZM or TZT structure in the minimal models.
\begin{table}[t]
\centering
\caption{
Gauged U(1) flavor symmetries which realize the TZM or TZT structures and their structural patterns.
The notations for the structural patterns of the TZM and TZT structures~\cite{Frampton:2002yf,Kageyama:2002zw} are written in Tab.~\ref{tab:tzmindex}. 
The index $\mathbf{R}$ ($\boldsymbol\nu$) means that the corresponding symmetry realizes TZM (TZT) structure of the light neutrino mass matrix.
Numbers in parentheses represent the U(1)$_{Y'}$ charges of the U(1)$_{Y'}$-breaking scalar singlets and doublets and ``$-$'' means that the corresponding U(1)$_{Y'}$ cannot realize the TZM or TZT structure.
}
\label{tab:tzmlist}
\vspace{5pt}
\begin{tabular}{lllll} \hline 
   & $Y'$ & Structural pattern \\ \cline{3-5}
   & & Singlet & Doublet \\ \hline 
   $e \mu \tau$ & $L_e-L_\mu$ & $\mathbf{E_1^R} (+1)$ & $\mathbf{A_2^{\boldsymbol\nu}} (+1)$ & $\mathbf{D_1^{\boldsymbol\nu}} (-1)$ \\ 
   & $L_\mu-L_\tau$ & $\mathbf{C^R} (+1)$ & $\mathbf{B_3^{\boldsymbol\nu}} (+1)$ & $\mathbf{B_4^{\boldsymbol\nu}} (-1)$ \\ 
   & $L_e-L_\tau$ & $\mathbf{E_2^R} (+1)$ & $\mathbf{A_1^{\boldsymbol\nu}} (+1)$ & $\mathbf{D_2^{\boldsymbol\nu}} (-1)$ \\ 
   $B e \mu \tau$ & $B-3L_e-L_\mu+L_\tau$ & $\mathbf{A_1^R} (+2)$ & $-$ & $-$ \\ 
   & $B-3L_e+L_\mu-L_\tau$ & $\mathbf{A_2^R} (+2)$ & $-$ & $-$ \\ 
   & $B-L_e-3L_\mu+L_\tau$ & $\mathbf{B_3^R} (+2)$ & $-$ & $-$ \\
   & $B-L_e+L_\mu-3L_\tau$ & $\mathbf{B_4^R} (+2)$ & $-$ & $-$ \\ 
   & $B+L_e-3L_\mu-L_\tau$ & $\mathbf{D_1^R} (+2)$ & $-$ & $-$ \\ 
   & $B+L_e-L_\mu-3L_\tau$ & $\mathbf{D_2^R} (+2)$ & $-$ & $-$ \\ 
   & $B-3L_e$ & $\mathbf{F_1^R} (+6)$ & $-$ & $-$ \\
   & $B-\frac{3}{2}L_\mu-\frac{3}{2}L_\tau$ & $\mathbf{F_1^R} (+3)$ & $-$ & $-$  \\ 
   & $B-3L_\mu$ & $\mathbf{F_2^R} (+6)$ & $-$ & $-$ \\
   & $B-\frac{3}{2}L_e-\frac{3}{2}L_\tau$ & $\mathbf{F_2^R} (+3)$ & $-$ & $-$ \\ 
   & $B-3L_\tau$ & $\mathbf{F_3^R} (+6)$ & $-$ & $-$ \\
   & $B-\frac{3}{2}L_e-\frac{3}{2}L_\mu$ & $\mathbf{F_3^R} (+3)$ & $-$ & $-$  \\ \hline
\end{tabular}
\label{tab:str-pattern}
\end{table}
The structural patterns in Tab.~\ref{tab:tzmlist} are summarized in Tab.~\ref{tab:tzmindex} and we follow the notation in Refs.~\cite{Frampton:2002yf,Kageyama:2002zw}.
\begin{table}[h]
\centering
\caption{
The list of the structural patterns of the two-zero minor and texture structure. We follow the notation in~\cite{Frampton:2002yf,Kageyama:2002zw}.
}
\label{tab:tzmindex}
\vspace{5pt}
\begin{tabular}{|c|cccccc:} \hline \hline
structure index & $\mathbf{A_1}$ & $\mathbf{A_2}$ & $\mathbf{B_3}$ & $\mathbf{B_4}$ & $\mathbf{C}$ & $\mathbf{D_1}$  \\ \hline
pattern & $\begin{pmatrix} 0 & 0 & * \\ 0 & * & * \\ * & * & * \end{pmatrix}$ & $\begin{pmatrix} 0 & * & 0 \\ * & * & * \\ 0 & * & * \end{pmatrix}$ & $\begin{pmatrix} * & 0 & * \\ 0 & 0 & * \\ * & * & * \end{pmatrix}$ & $\begin{pmatrix} * & * & 0 \\ * & * & * \\ 0 & * & 0 \end{pmatrix}$ & $\begin{pmatrix} * & * & * \\ * & 0 & * \\ * & * & 0 \end{pmatrix}$ & $\begin{pmatrix} * & * & * \\ * & 0 & 0 \\ * & 0 & * \end{pmatrix}$  \\ \hline 
\multicolumn{7}{c}{} \\ \cline{2-7} \noalign{\vskip-1mm}
\multicolumn{1}{c}{} & \multicolumn{1}{:c}{$\mathbf{D_2}$} & $\mathbf{E_1}$ & $\mathbf{E_2}$ & $\mathbf{F_1}$ & $\mathbf{F_2}$ & \multicolumn{1}{c|}{$\mathbf{F_3}$} \\ \cline{2-7}
\multicolumn{1}{c}{} & \multicolumn{1}{:c}{$\begin{pmatrix} * & * & * \\ * & * & 0 \\ * & 0 & 0 \end{pmatrix}$} & $\begin{pmatrix} 0 & * & * \\ * & 0 & * \\ * & * & * \end{pmatrix}$ & $\begin{pmatrix} 0 & * & * \\ * & * & * \\ * & * & 0 \end{pmatrix}$ & $\begin{pmatrix} * & 0 & 0 \\ 0 & * & * \\ 0 & * & * \end{pmatrix}$ & $\begin{pmatrix} * & 0 & * \\ 0 & * & 0 \\ * & 0 & * \end{pmatrix}$ & \multicolumn{1}{c|}{$\begin{pmatrix} * & * & 0 \\ * & * & 0 \\ 0 & 0 & * \end{pmatrix}$} \\  \cline{2-7} \noalign{\vskip1mm} \cline{2-7}
\end{tabular}
\end{table}
As we see in Tab.~\ref{tab:tzmlist}, there are only 15 patterns, and only the minimal gauged U(1)$_{L_\alpha-L_\beta}$ models with a doublet scalar can realize the TZT structures.
These doublet cases have been discussed in detail in Ref.~\cite{Asai:2018ocx}, and so we explain them briefly and show only the results for completeness in this paper.
Moreover, we can use the same method as Ref.~\cite{Asai:2017ryy,Asai:2018ocx} to analyze the neutrino mass matrices in the cases of general U(1)$_{Y'}$s because the difference between the U(1)$_{L_\mu-L_\tau}$ and the other U(1)$_{Y'}$ cases are only the positions of the zero components.

\subsection{Minimal gauged U(1) models}
\label{subsec:models}

Here, using U(1)$_{B+L_e-3L_\mu-L_\tau}$ case, we show the realization of the TZM structure in the minimal gauged U(1)$_{Y'}$ model with a U(1)$_{Y'}$-breaking singlet scalar.
The interaction terms in the leptonic sector are then given by
\begin{align}
 \Delta {\cal L} = 
 & - y_e e_R^c L_e H^\dag - y_\mu \mu_R^c L_\mu H^\dag - y_\tau \tau_R^c L_\tau H^\dag \nonumber \\
&-\lambda_e N_e^c (L_e \cdot H)
-\lambda_\mu N_\mu^c (L_\mu \cdot H)
-\lambda_\tau N_\tau^c (L_\tau \cdot H) \nonumber \\
& - M_{e \tau} N_e^c N_\tau^c 
- \frac{1}{2} \lambda_{ee} \sigma N_e^c N_e^c
- \lambda_{e\mu} \sigma^* N_e^c N_\mu^c
- \frac{1}{2} \lambda_{\tau \tau} \sigma^* N_\tau^c N_\tau^c +\text{h.c.} ~,
\label{eq:lag}
\end{align}
where the dots indicate the contraction of the SU(2)$_L$ indices. 
After the Higgs field $H$ and the singlet scalar $\sigma$ acquire VEVs\footnote{We can always take the VEV of $\sigma$ to be real by using U(1)$_{B+L_e-3L_\mu-L_\tau}$ transformations. } $\braket{H} = v/\sqrt{2}$ and $\braket{\sigma}$, the Dirac, Majorana and charged lepton mass matrices are obtained as follows:
\begin{equation}
 {\cal M}_D = \frac{v}{\sqrt{2}}
\begin{pmatrix}
 \lambda_e & 0& 0\\
 0 & \lambda_\mu & 0 \\
 0 & 0 & \lambda_\tau 
\end{pmatrix}
~,~
{\cal M}_R =
\begin{pmatrix}
 \lambda_{ee} \braket{\sigma} & \lambda_{e\mu} \braket{\sigma} & M_{e\tau}  \\
 \lambda_{e\mu} \braket{\sigma} & 0 & 0 \\
M_{e\tau} & 0 & \lambda_{\tau \tau} \braket{\sigma}
\end{pmatrix}
~,~
 {\cal M}_\ell = \frac{v}{\sqrt{2}}
\begin{pmatrix}
 y_e & 0& 0\\
 0 & y_\mu & 0 \\
 0 & 0 & y_\tau 
\end{pmatrix}
~.
\end{equation}
Throughout this work, we assume that the non-zero components in the Majorana mass matrix $\mathcal{M}_R$ are much larger than those in the Dirac matrix $\mathcal{M}_D$ so that the mass matrix of the active neutrinos is given by the seesaw formula~\cite{Minkowski:1977sc, Yanagida:1979as, GellMann:1980vs,Mohapatra:1979ia} 
\begin{equation}
 {\cal M}_{\nu_L} \simeq - {\cal M}_D {\cal M}_R^{-1} {\cal M}_D^T ~.
\label{eq:mnul}
\end{equation}
We note that any size of the VEV of the singlet scalar field $\sigma$ can be allowed as long as the Majorana masses are much larger than the Dirac masses.

The minimal gauged U(1)$_{Y'}$ models are constrained by many kinds of the experiments relevant to the U(1)$_{Y'}$ gauge boson $Z'$~\cite{Wise:2018rnb,Bauer:2018onh,Chun:2018ibr}.
For example, $Z'$ with relatively large gauge coupling like $\gtrsim \mathcal{O}(10^{-3})$ is constrained by the $e^+ e^-$ colliders, such as BABAR~\cite{Lees:2017lec}, Belle~\cite{Won:2016pjz,DePietro:2018sgg}, KLOE~\cite{Babusci:2012cr,Anastasi:2016ktq,Anastasi:2018azp}, etc.
When $Z'$ couples to electrons, these experiments give strict constraints.
Otherwise, when $Z'$ couples to muons, for example, the BABAR experiment search for $e \bar{e} \to \mu \bar{\mu} Z'$, $Z' \to \mu \bar{\mu}$ constrains $Z'$ in around GeV mass range~\cite{TheBABAR:2016rlg}.  
Other than that, there is the Borexino experiment~\cite{Bellini:2011rx,Harnik:2012ni}, which gives a bound on the $\nu \mathchar`-e$ interactions and the tests of neutrino trident production~\cite{Altmannshofer:2014cfa,Geiregat:1990gz,Mishra:1991bv,Altmannshofer:2014pba}.
For relatively light mass ($m_{Z'} \lesssim 100$ MeV) and small coupling constant ($g_{Z'} \lesssim 10^{-3}$), the minimal models are constrained by the electron beam-dump experiments~\cite{Bjorken:2009mm,Andreas:2012mt}, such as SLAC E137~\cite{Bjorken:1988as}, SLAC E141~\cite{Riordan:1987aw}, Fermilab E774~\cite{Bross:1989mp}, Orsay~\cite{Davier:1989wz}, etc, and the proton beam-dump experiments, such as CHARM~\cite{Bergsma:1985qz,Gninenko:2012eq}, U70/Nu-Cal~\cite{Blumlein:2011mv,Blumlein:2013cua}, etc.

The doublet case realizes the TZT structure.
As we have discussed in Ref.~\cite{Asai:2018ocx}, the difference between the singlet and doublet cases is that a doublet scalar with U(1)$_{Y'}$ charge $+1$ realizes different TZT structure from one with U(1)$_{Y'}$ charge $-1$~\footnote{In the singlet case, the exchange of U(1)$_{Y'}$ charge $+1$ for $-1$ just corresponds to that of $\sigma$ for $\sigma^*$ and the realized neutrino mass structure does not change.}.
Furthermore, the U(1)$_{Y'}$-breaking scale should be lower than the electroweak scale, and only the U(1)$_{L_\mu-L_\tau}$ avoids the experimental constraints relevant to $Z'$.

\section{Vanishing conditions and neutrino parameters} 
\label{sec:vanishing}
In this section, following the analyses of Refs.~\cite{Asai:2017ryy,Asai:2018ocx}, we analyze the conditions from two-zero minor (texture) structure in the active neutrino mass matrix and see the two conditional equations which the low-energy neutrino parameters should satisfy. By solving these equations, we show that four unknown parameters can be written as functions of the rest of the parameters.
As we mentioned before, we explain how to analyze neutrino mass matrix, using the U(1)$_{B+L_e-3L_\mu-L_\tau}$ case, and this analyses can be applied to the other U(1)$_{Y'}$s.

\subsection{Vanishing conditions in the neutrino mass matrix}
\label{subsec:vanishing}
As we said in the previous section, we assume that the Majorana masses are much larger than the Dirac masses so that the mass matrix of the active neutrinos is given by the seesaw formula (Eq.~\eqref{eq:mnul}).
On the other hand, the charged mass matrix $\mathcal{M}_\ell$ is diagonal in this model, and thus we can obtain the mass eigenvalues of the light neutrinos, diagonalizing this matrix by a unitary matrix $U_{\text{PMNS}}$:
\begin{equation}
 U_{\text{PMNS}}^T {\cal M}_{\nu_L} U_{\text{PMNS}} =\text{diag}(m_1, m_2, m_3) ~,
\label{eq:diagmnul}
\end{equation}
where $U_{\text{PMNS}}$ is the so-called Pontecorvo-Maki-Nakagawa-Sakata (PMNS) mixing matrix~\cite{Pontecorvo:1967fh,Pontecorvo:1957cp,Pontecorvo:1957qd,Maki:1962mu}.
The PMNS matrix is parametrized as 
\begin{equation}
 U_{\text{PMNS}} = 
\begin{pmatrix}
 c_{12} c_{13} & s_{12} c_{13} & s_{13} e^{-i\delta} \\
 -s_{12} c_{23} -c_{12} s_{23} s_{13} e^{i\delta}
& c_{12} c_{23} -s_{12} s_{23} s_{13} e^{i\delta}
& s_{23} c_{13}\\
s_{12} s_{23} -c_{12} c_{23} s_{13} e^{i\delta}
& -c_{12} s_{23} -s_{12} c_{23} s_{13} e^{i\delta}
& c_{23} c_{13}
\end{pmatrix}
\begin{pmatrix}
 1 & & \\
 & e^{i\frac{\alpha_{2}}{2}} & \\
 & & e^{i\frac{\alpha_{3}}{2}}
\end{pmatrix}
~,
\end{equation}
where $c_{ij} \equiv \cos \theta_{ij}$ and $s_{ij} \equiv \sin \theta_{ij}$ for $\theta_{ij} = [0, \pi/2]$, $\delta, \alpha_2, \alpha_3 = [0, 2\pi]$, and we have ordered $m_1<m_2$ without loss of generality. 
We follow the convention of the Particle Data Group~\cite{Olive:2016xmw}, where $m_2^2-m_1^2\ll |m_3^2-m_1^2|$ and $m_1<m_2<m_3$ (Normal Ordering, NO) or $m_3<m_1<m_2$ (Inverted Ordering, IO). 
In Tab.~\ref{tab:input}, we show the best fit values for the neutrino oscillation parameters and $1\sigma \  (3\sigma)$ ranges based on the global analysis of neutrino data from the \texttt{NuFIT v4.0} result with the Super-Kamiokande atmospheric data ~\cite{nufit,Esteban:2018azc}.
Here, we take $\ell=1$ for NO and $\ell=2$ for IO in $\Delta m_{3\ell}^2$~\cite{Esteban:2016qun}.

\begin{table}[t]
 \begin{center}
\caption{Values for the neutrino oscillation parameters we use
  in this paper. We take them from the \texttt{NuFIT v4.0} result
  with the Super-Kamiokande atmospheric data
 ~\cite{nufit, Esteban:2018azc}.  }
\label{tab:input}
\vspace{5pt}
\begin{tabular}{l|cc|cc}
\hline
\hline
 & \multicolumn{2}{c|}{Normal Ordering} &
 \multicolumn{2}{c}{Inverted Ordering} \\ 
\cline{2-5}
Parameter  & Best fit $\pm 1\sigma$& 3$\sigma$ range \qquad& Best fit
 $\pm 1\sigma$& 3$\sigma$ range \qquad  \\
\hline
$\sin^2 \theta_{12}$ & $0.310^{+0.013}_{-0.012}$ & 0.275--0.350 &
 $0.310^{+0.013}_{-0.012}$ & 0.275--0.350 \\
$\sin^2 \theta_{23}$ & $0.582^{+0.015}_{-0.019}$ & 0.428--0.624 &
 $0.582^{+0.015}_{-0.018}$ & 0.433--0.623 \\
$\sin^2 \theta_{13}$ & $0.02240^{+0.00065}_{-0.00066}$ & 0.02044--0.02437 & $0.02263^{+0.00065}_{-0.00066}$
 & 0.02067--0.02461 \\
$\Delta m^2_{21}/10^{-5}~\text{eV}^2$ & $7.39^{+0.21}_{-0.20}$ &
	 6.79--8.01 & $7.39^{+0.21}_{-0.20}$ & 6.79--8.01  \\
$\Delta m^2_{3\ell}/10^{-3}~\text{eV}^2$ & $2.525^{+0.033}_{-0.031}$ &
	 $2.431$--$2.622$ 
	 &$-2.512^{+0.034}_{-0.031}$ & $-$(2.606--$2.413$) \\
$\delta ~[{}^\circ]$ & $217^{+40}_{-28}$ & 135--366 & $280^{+25}_{-28}$ & 196--351\\
\hline
\hline
\end{tabular}
 \end{center}
\end{table}

In this paper, we consider only the $m_i \neq 0$ cases like Refs.~\cite{Asai:2017ryy,Asai:2018ocx}. 
This is because if $m_i = 0$ ($ i = 1$ or $3$), the mass matrix for the light neutrinos $\mathcal{M}_{\nu_L}$ is block-diagonal, and we cannot have desired mixing angles. 
From Eqs.~\eqref{eq:mnul} and \eqref{eq:diagmnul}, the following relation is given:
\begin{align}
 {\cal M}_{\nu_L}^{-1} = - ({\cal M}_D^{-1})^T {\cal M}_R {\cal M}_D^{-1}
 = U_{\text{PMNS}}  \text{diag}(m_1^{-1}, m_2^{-1}, m_3^{-1}) U_{\text{PMNS}}^T~.
\end{align}
$\mathcal{M}_D$ is diagonal and $(\mu,\mu)$ and $(\mu,\tau)$ components in $\mathcal{M}_R$ vanish, and so these components of the right-hand side also should vanish.
Then the minimal gauged U(1)$_{B+L_e-3L_\mu-L_\tau}$ model is one of the concrete realizations of the two-zero-minor structure. 
From these vanishing conditions, the following two equations are given:
\begin{align}
 \frac{1}{m_1}V_{\mu 1}^2 + \frac{1}{m_2}V_{\mu 2}^2\,e^{i\alpha_2}
+ \frac{1}{m_3}V_{\mu 3}^2\,e^{i\alpha_3} &= 0 ~,
\label{eq:vanmu}\\[3pt]
 \frac{1}{m_1}V_{\mu 1}V_{\tau 1} + \frac{1}{m_2}V_{\mu 2}V_{\tau 2}\,e^{i\alpha_2}
+ \frac{1}{m_3}V_{\mu 3}V_{\tau 3}\,e^{i\alpha_3} &= 0 ~,
\label{eq:vantau}
\end{align}
where the unitary matrix $V$ is defined by $U_{\text{PMNS}} = V\cdot \text{diag}(1, e^{i\alpha_2/2}, e^{i\alpha_3/2})$. 
We notice that neither the VEV of the U(1)$_{B+L_e-3L_\mu-L_\tau}$-symmetry breaking singlet scalar $\langle \sigma \rangle$ nor Majorana masses $M_{e\tau}$ appear in these conditions explicitly. 
This is the most important point in this analysis.
This is because the following discussions and results based on the above conditions are independent of the U(1)$_{B+L_e-3L_\mu-L_\tau}$-symmetry breaking scale and the Majorana mass scale. 
Needless to say, to explain the lightness of the active neutrinos, we use the seesaw mechanism in Eq.~\eqref{eq:mnul}, and so these scales lie around much greater than the scale of the light neutrinos. 
Moreover, it is shown in Ref.~\cite{Asai:2017ryy} that the two-zero minor structure is preserved under the renormalization group flow when the charged lepton mass matrix is diagonal.  

In the doublet cases, the TZT structure of the neutrino mass matrix gives the following vanishing conditions:
\begin{align}
\label{eq:mnul-d}
   \mathcal{M}_{\nu_L} = - \mathcal{M}_D \mathcal{M}_R^{-1} \mathcal{M}_D^T = U_{\text{PMNS}}^* \text{diag}(m_1,m_2,m_3) U_{\text{PMNS}}^\dag~.
\end{align}
In the case of the minimal gauged U(1)$_{L_\mu-L_\tau}$ model with a doublet scalar with U(1)$_{L_\mu-L_\tau}$ charge $+1$, both sides of Eq.~\eqref{eq:mnul-d} have zeros in the $(e,\mu)$ and $(\mu,\mu)$ components.
Then, we obtain the vanishing conditions from Eq.~\eqref{eq:mnul-d} and can analyze these in the same way as the singlet case.

\subsection{Analyses of vanishing conditions}
\label{subsec:analyses}
\begin{figure}[htpb]
\centering
\includegraphics[clip, width = 0.5 \textwidth]{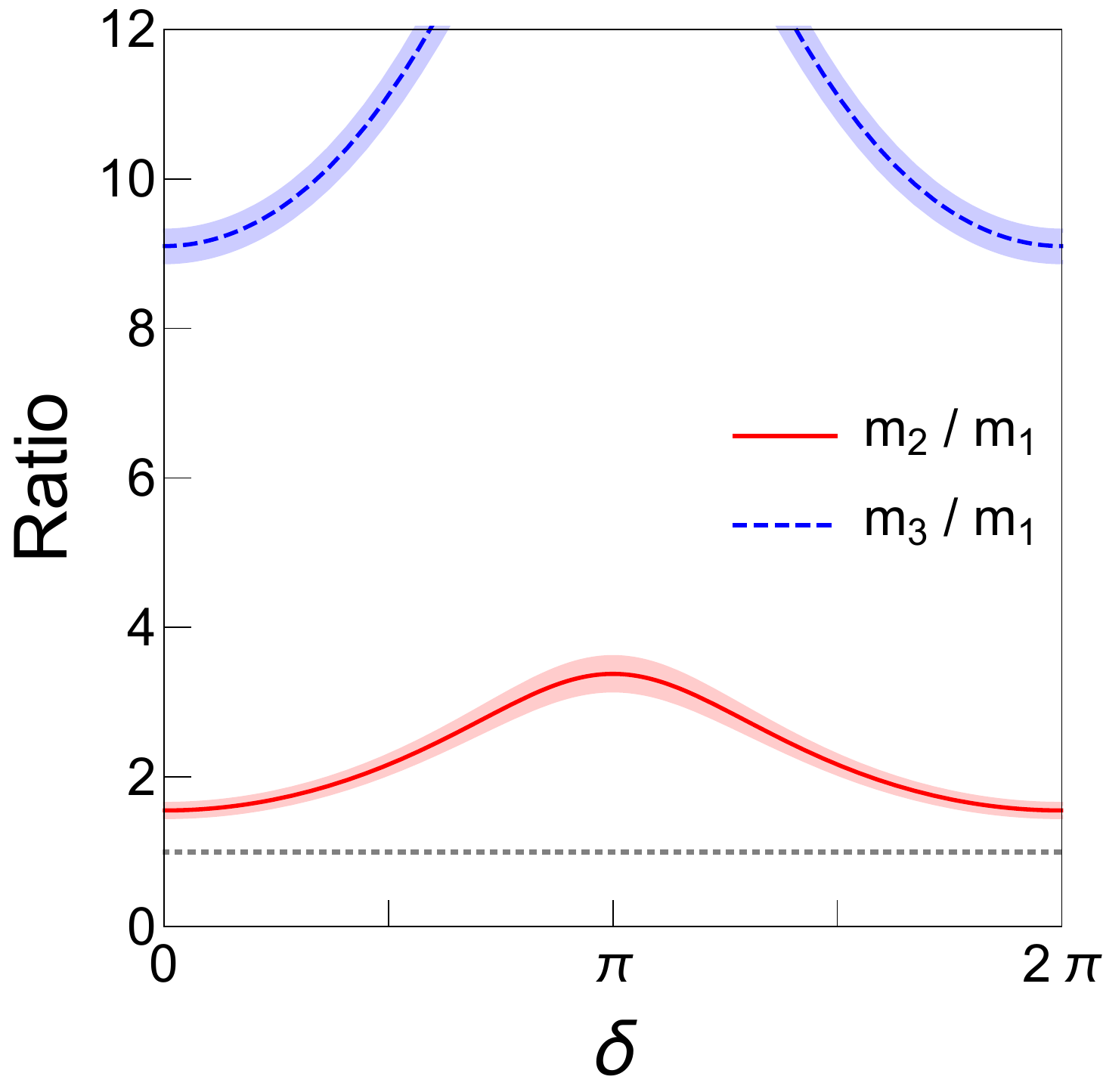}
\caption{
The mass ratios $m_2/m_1$ and $m_3/m_1$ as functions of the Dirac CP phase $\delta$ in the minimal gauged U(1)$_{B+L_e-3L_\mu-L_\tau}$ model with a singlet scalar. 
The bands show uncertainty coming from the 1$\sigma$ error in $\theta_{12}$, which gives dominant contributions. 
The thin gray dotted line corresponds to $m_{2,3}/m_1 = 1$. 
For input parameters of the neutrino mixing angles, we use the NO values in Tab.~\ref{tab:input}.
}
\label{fig:ratio}
\end{figure}

Eqs.~\eqref{eq:vanmu} and \eqref{eq:vantau} are two complex equations. 
Therefore, by solving these conditional equations, we can obtain four real parameters, namely the Dirac CP phase $\delta$, Majorana CP phases $\alpha_{2,3}$, and mass eigenvalue of the active neutrino $m_1$, as functions of the neutrino oscillation parameters $\theta_{12}$, $\theta_{23}$, $\theta_{13}$, $\Delta m^2_{21}$, and $\Delta m^2_{3\ell}$. 
We analyze these conditions, following Refs.~\cite{Asai:2017ryy,Asai:2018ocx}. From Eqs.~\eqref{eq:vanmu} and \eqref{eq:vantau}, we obtain
\begin{align}
 e^{i\alpha_2} &=\frac{m_2}{m_1} R_2 (\delta) ~,
\qquad
 e^{i\alpha_3} =\frac{m_3}{m_1} R_3 (\delta) ~,
\label{eq:alp23}
\end{align}
with
\begin{align}
\label{eq:r2r3}
 R_2 \equiv \frac{V_{\mu 1} V^*_{e2}}{V_{\mu 2} V^*_{e1}} ~,
 \qquad
 R_3 \equiv \frac{V_{\mu 1} V^*_{e3}}{V_{\mu 3} V^*_{e1}} ~,
\end{align}
where we have used $\widetilde{V}^T = V^{-1}$ and $\textrm{det}V=1$ with $\widetilde{V}$ being the cofactor matrix of $V$.\footnote{From $\tilde{V}^T = V^{-1}$, we obtain $V^* = \tilde{V}$ , and for instance, the following relation holds: $V^*_{e1} = V_{\mu 2} V_{\tau 3} - V_{\mu 3} V_{\tau 2}$.}
In Appendix~\ref{app:r2r3}, we give the explicit expressions for $R_2$ and $R_3$ in terms of neutrino oscillation parameters.
By taking absolute values of Eqs.~\eqref{eq:alp23}, we obtain the mass ratios $m_2/m_1$, $m_3/m_1$ as functions of the Dirac CP phase:
\begin{equation}
 \frac{m_2}{m_1} = \frac{1}{|R_2(\delta)|} ~,\qquad
 \frac{m_3}{m_1} = \frac{1}{|R_3(\delta)|} ~.
\label{eq:m2m3}
\end{equation}
In Fig.~\ref{fig:ratio}, we show the mass ratios $m_2/m_1$ and $m_3/m_1$ against the Dirac CP phase $\delta$. 
The bands show uncertainty coming from the 1$\sigma$ error in $\theta_{12}$, which gives dominant contributions. 
For input parameters of the neutrino mixing angles, we use the NO values in Tab.~\ref{tab:input}.
We find that the NO: $m_1 < m_2 < m_3$ is realized in $0 \leq \delta \leq 2\pi$ from this figure, while the IO is not.

We have calculated the mass ratios in the other U(1)$_{Y'}$ extended models in the same way as the above example, and then have found that only five U(1)$_{Y'}$ symmetries realize the NO and two do the IO\footnote{This result that five U(1)$_{Y'}$ symmetries realize the NO or the IO is consistent with Refs.~\cite{Araki:2012ip,Heeck:2014sna}.}. 
We show the summary of the consistency between the neutrino oscillation parameters and mass ordering in Tab.~\ref{tab:MO}.
\begin{table}[t]
\centering
\caption{
The consistency between the neutrino oscillation parameters and the mass ordering. 
``$\checkmark$'' represents the compatibility of the TZM structure with NO or IO when the best fit value of $\theta_{23}$ is used, and ``$\times$'' does the opposite. 
``$\triangle$'' in the $\mathbf{B_3^R}$ and $\mathbf{B_4^R}$ cases mean that the NO or IO cannot be realized when we use the best fit value of $\theta_{23}$, but they are realized when we use the smallest value in the $3\sigma$ region of $\theta_{23}$.
The subscript $\mathbf{R}$ means that the neutrino mass matrix has the two-zero minor structure.
}
\label{tab:MO}
\vspace{5pt}
\begin{tabular}{ccccccccccccc} \hline \hline
structure index & $\mathbf{A_1^R}$ & $\mathbf{A_2^R}$ & $\mathbf{B_3^R}$ & $\mathbf{B_4^R}$ & $\mathbf{C^R}$ & $\mathbf{D_1^R}$ & $\mathbf{D_2^R}$ & $\mathbf{E_1^R}$ & $\mathbf{E_2^R}$ & $\mathbf{F_1^R}$ & $\mathbf{F_2^R}$ & $\mathbf{F_3^R}$ \\ \hline
NO & $\times$ & $\times$ & $\checkmark$ & $\triangle$ & $\checkmark$ & $\checkmark$ & $\checkmark$ & $\times$ & $\times$ & $\times$ & $\times$ & $\times$ \\
IO & $\times$ & $\times$ & $\triangle$ & $\checkmark$ & $\times$ & $\times$ & $\times$ & $\times$ & $\times$ & $\times$ & $\times$ & $\times$ \\ \hline \hline
\end{tabular}
\end{table}
``$\checkmark$'' represents the compatibility of the TZM structure with the NO or IO when the best fit value of $\theta_{23}$ is used, and ``$\times$'' does the opposite. 
``$\triangle$'' in the $\mathbf{B_3^R}$ and $\mathbf{B_4^R}$ cases mean that the NO or IO cannot be realized when we use the best fit value of $\theta_{23}$, but they are realized when we use the smallest value in the $3\sigma$ region of $\theta_{23}$.
From this result, we discuss only the U(1)$_{L_\mu-L_\tau} [\mathbf{C^R}]$, U(1)$_{B-L_e-3L_\mu+L_\tau} [\mathbf{B_3^R}]$, U(1)$_{B-L_e+L_\mu-3L_\tau} [\mathbf{B_4^R}]$, U(1)$_{B+L_e-3L_\mu-L_\tau} [\mathbf{D_1^R}]$ and U(1)$_{B+L_e-L_\mu-3L_\tau} [\mathbf{D_2^R}]$ cases, henceforth.

Returning to the U(1)$_{B+L_e-3L_\mu-L_\tau}$ [$\mathbf{D_1^R}$] case, from the above discussion, we fix the mass ordering to the NO and determine $m_1$ and $\delta$ as functions of neutrino oscillation parameters. 
From Eqs.~\eqref{eq:m2m3}, an equation for $\cos \delta$ is obtained by
\begin{align}
   \epsilon |R_2(\delta)|^2 (1-|R_3(\delta)|^2) = |R_3(\delta)|^2 (1-|R_2(\delta)|^2)
\label{eq:cos-eq}
\end{align}
where $\epsilon \equiv \Delta m_{21}^2/\Delta m_{31}^2$, and it leads $\delta$ as functions of neutrino oscillation parameters. 
We give an explicit expression of Eq.~\eqref{eq:cos-eq} in Eq.~\eqref{eq:D1eqofdelta} in Appendix~\ref{sec:cub}.
Substituting the best fit values of the mixing angles $\theta_{ij}$ in Tab.~\ref{tab:input} into Eq.~\eqref{eq:D1eqofdelta} and solving it numerically, we obtain $\cos \delta \simeq 0.102$ for the best fit values of the neutrino oscillation parameters, which means $\delta = 0.468\pi$ or $1.532\pi$. 

Substituting this result into Eqs.~\eqref{eq:m2m3}, we also obtain $m_1$ as a function of neutrino oscillation parameters.
An explicit expression for $m_1$ is given in Appendix~\ref{app:m1}.
Moreover, comparing the phases of $R_{2,3}$ in Eqs.~\eqref{eq:alp23} respectively, we can obtain the Majorana CP phases:
\begin{align}
\label{eq:MajoranaCPphase}
   \alpha_2 = \text{Arg}\left[ \frac{m_2}{m_1} R_2(\delta) \right]~,\ \ \ \alpha_3 = \text{Arg}\left[ \frac{m_3}{m_1} R_3(\delta) \right]~.
\end{align}
It is shown in Ref.~\cite{Asai:2017ryy} that $|R_2(\delta)|$ and $|R_3(\delta)|$ do not change under $\delta \to -\delta$, and so $m_1$ also does not change whether we choose $\delta$ or $-\delta$. 
Furthermore, we have $\alpha_{2,3} (-\delta) = -\alpha_{2,3}(\delta)$. 
These discussions are applicable to the other U(1)$_{Y'}$ symmetries and we obtain the same conclusions.

\vspace{-3mm}
\section{Predictions for the neutrino parameters}
\label{sec:result}
In the previous section, we have shown that, by analyzing the two-zero minor and texture structures, we can obtain the neutrino mass, Dirac CP phase and Majorana CP phases as functions of the neutrino oscillation parameters, namely the three mixing angles and two squared mass differences. 
Here, we show the results of calculations with the errors in the neutrino oscillation parameters. 
For input values, we use the values given in Tab.~\ref{tab:input}~\cite{nufit,Esteban:2018azc}. 
Moreover, we show the result of the effective neutrino mass relevant to the neutrinoless double beta decay.
The results in the U(1)$_{L_\mu-L_\tau}$ case are not new ones, but ones of our previous work~\cite{Asai:2018ocx}.
However, we show them in this paper for completeness.

\subsection{Neutrino masses}
\label{subsec:result-mass}
\begin{figure}[htpb]
\begin{subfigure}[b]{0.32\linewidth}
\centering
\includegraphics[width=5cm]{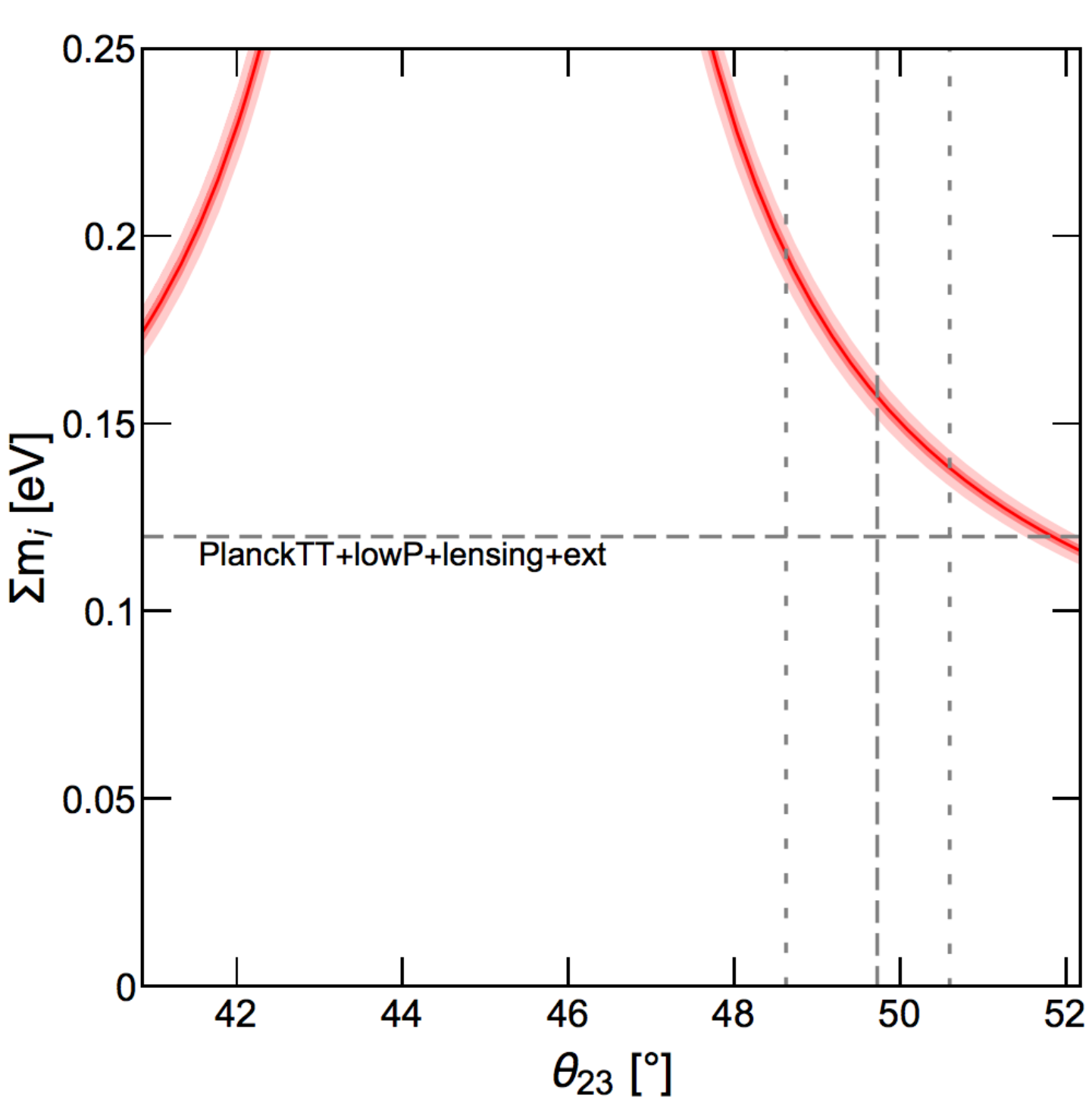}
\subcaption{$\mathbf{C^R}$ (NO)}
\label{fig:mass-CR-NO} 
\end{subfigure}
\begin{subfigure}[b]{0.32\linewidth}
\centering
\includegraphics[width=5cm]{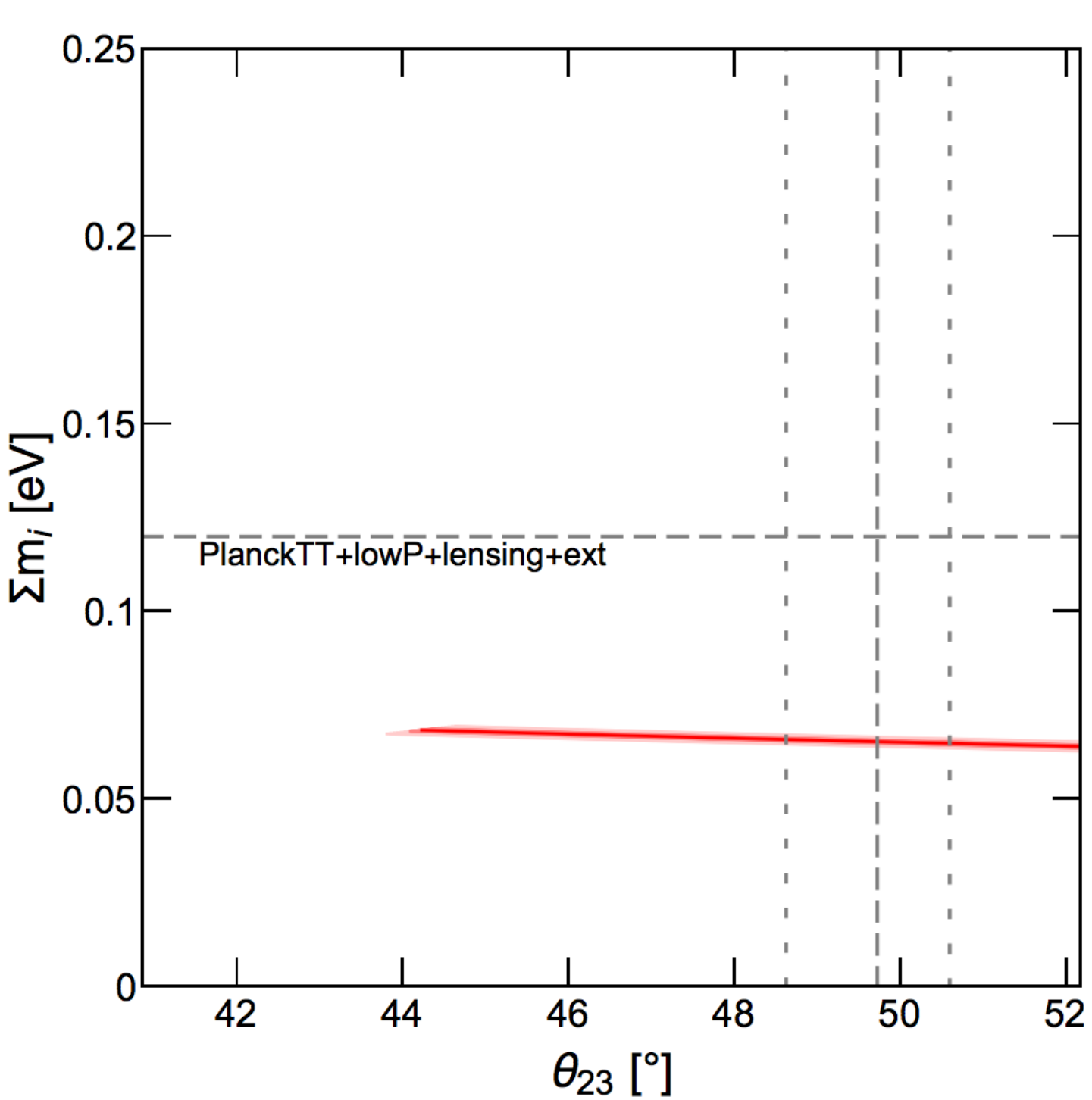}
\subcaption{$\mathbf{D_1^R}$ (NO)}
\label{fig:mass-D1R-NO}
\end{subfigure}
\begin{subfigure}[b]{0.32\linewidth}
\includegraphics[width=5cm]{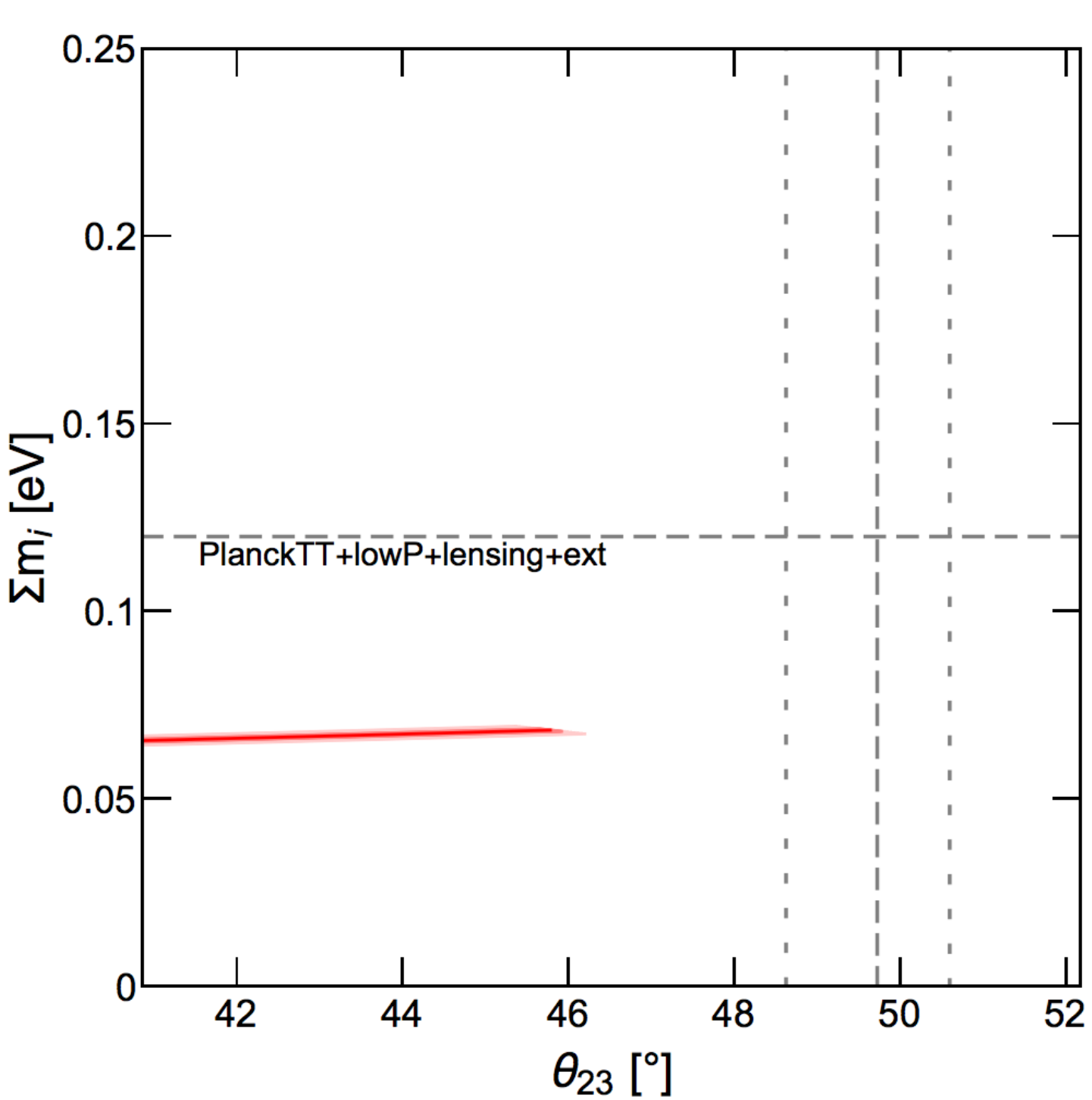}
\subcaption{$\mathbf{D_2^R}$ (NO)}
\label{fig:mass-D2R-NO}
\end{subfigure} \\
\begin{subfigure}[b]{0.5\linewidth}
\centering
\setcounter{subfigure}{3}
\includegraphics[width=5cm]{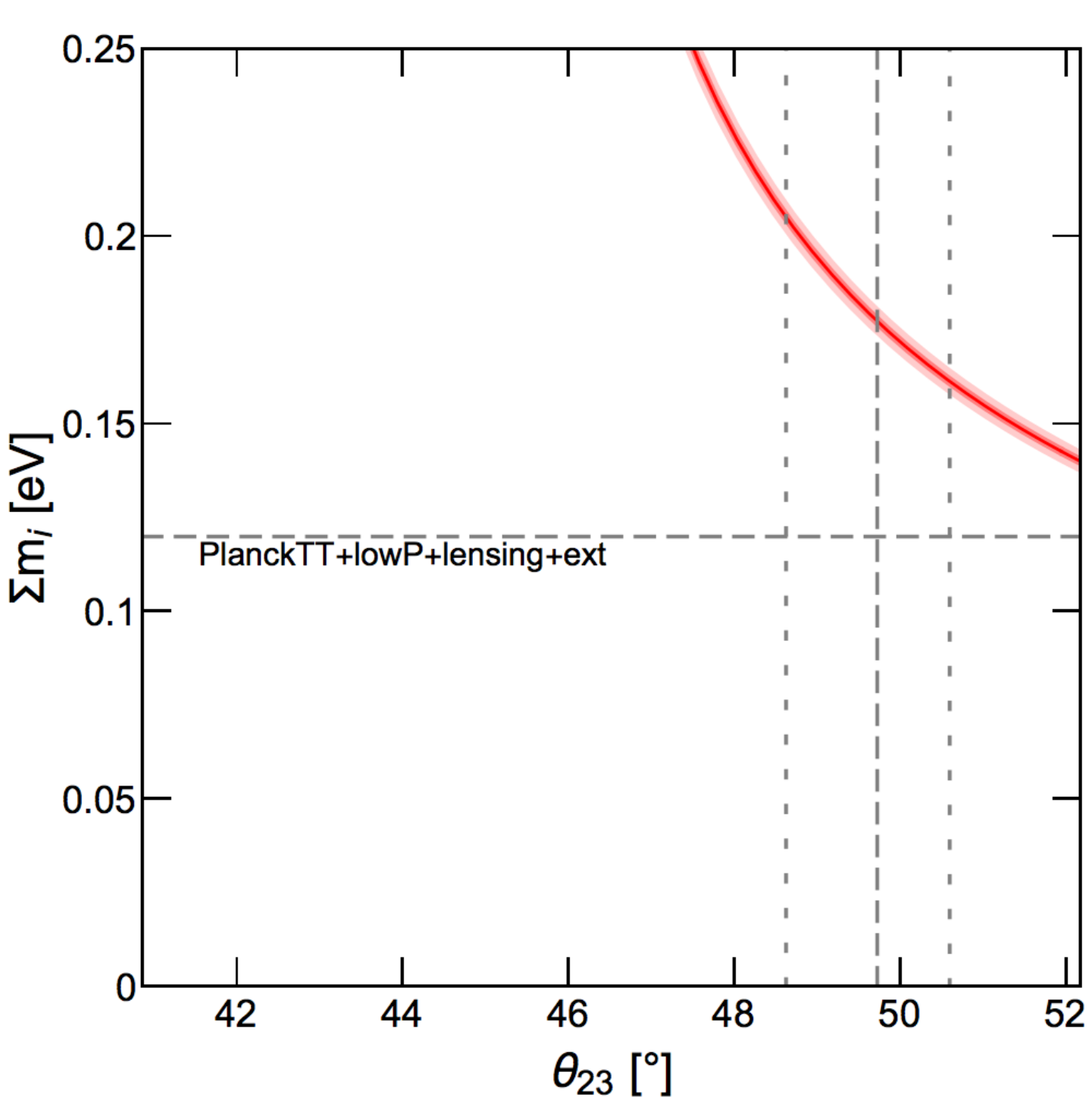}
\subcaption{$\mathbf{B_3^R}$ (NO)}
\label{fig:mass-B3R-NO}
\end{subfigure}
\begin{subfigure}[b]{0.4\linewidth}
\includegraphics[width=5cm]{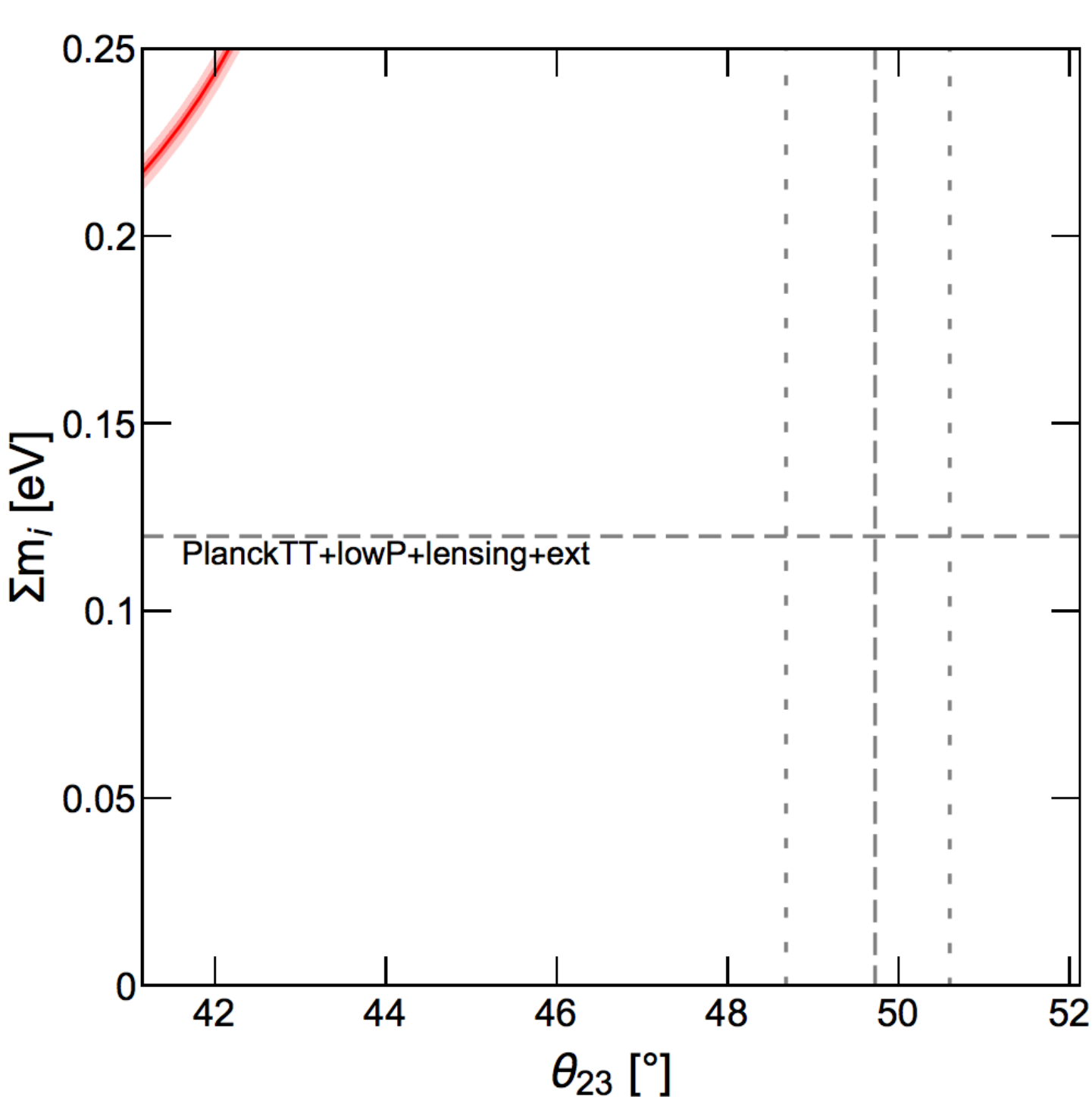}
\subcaption{$\mathbf{B_3^R}$ (IO)}
\label{fig:mass-B3R-IO}
\end{subfigure}
\begin{subfigure}[b]{0.5\linewidth}
\centering
\includegraphics[width=5cm]{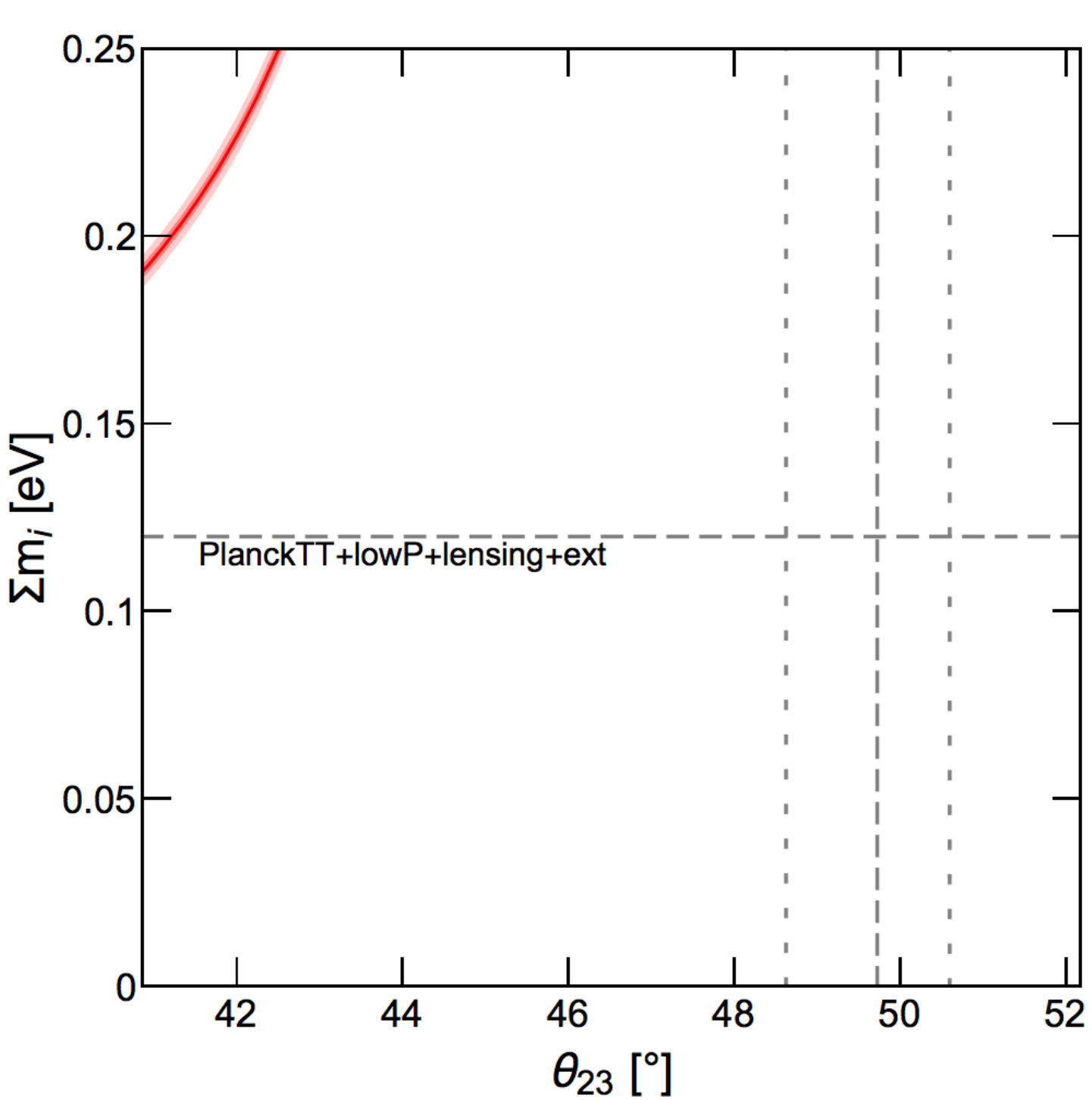}
\subcaption{$\mathbf{B_4^R}$ (NO)}
\label{fig:mass-B4R-NO}
\end{subfigure} \hspace{13mm}
\begin{subfigure}[b]{0.4\linewidth}
\includegraphics[width=5cm]{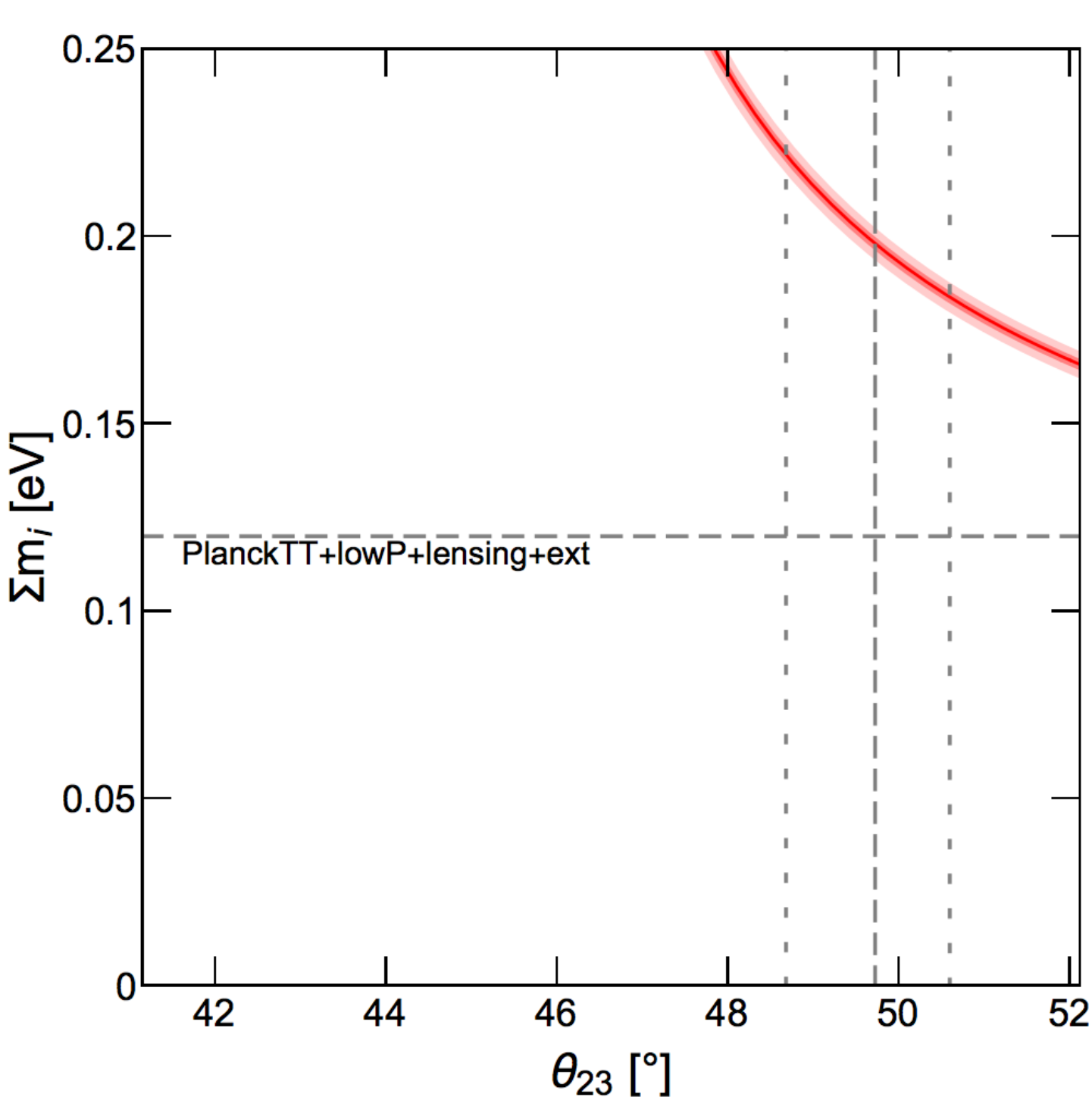}
\subcaption{$\mathbf{B_4^R}$ (IO)}
\label{fig:mass-B4R-IO}
\end{subfigure}
\caption{
The predictions for the sum of the neutrino masses in the singlet models.  
The red line shows the prediction as functions of $\theta_{23}$ and the dark (light) red bands show the uncertainty coming from the $1\sigma$ ($3\sigma$) errors of $\theta_{13}$ and $\Delta m^2$ in the $\mathbf{C^R}$ and the other structures, respectively. 
The vertical gray dashed line represents the best fit value of $\theta_{23}$ and the vertical gray dotted lines represent the $1\sigma$ region. 
The horizontal dashed gray line shows the present limit on the sum of the neutrino masses by the Planck experiment:~$\sum_i m_i < 0.12$~eV~(Planck TT+lowP+lensing+ext)~\cite{Aghanim:2018eyx}.
}
\label{fig:mass-sum-s}
\end{figure}

In the singlet cases, the U(1)$_{Y'}$ gauge symmetries which realize the NO or the IO are only five, as shown in Tab.~\ref{tab:MO}.
In Fig.~\ref{fig:mass-sum-s}, we plot the sum of the neutrino masses as functions of $\theta_{23}$ in the red line.
$\theta_{23}$ is varied in the $3\sigma$ region. 
Fig.~\ref{fig:mass-sum-s}~(\subref{fig:mass-CR-NO}) is the same as Fig.~1 in Ref.~\cite{Asai:2018ocx}.
The vertical gray dashed line represents the best fit value of $\theta_{23}$, and the vertical gray dotted lines (the plot range) indicate the $1\sigma$ ($3\sigma$) region.
The dark (light) red bands show the uncertainty coming from the $1\sigma$ ($3\sigma$) errors of $\theta_{13}$ and $\Delta m^2_{31}$ in the $\mathbf{C^R}$ and the other structures, respectively. 
The contribution to $\sum m_i$ from the errors of the other parameters are subdominant, and so we take them to be the best fit values.
The horizontal dashed line shows the present limit on the sum of the neutrino masses by the Planck experiment:~$\sum_i m_i < 0.12$~eV~(Planck TT+lowP+lensing+ext)~\cite{Aghanim:2018eyx}.
As we see in Fig.~\ref{fig:mass-sum-s}~(\subref{fig:mass-CR-NO}), in the $\mathbf{C^R}$ case, there is a strong tension between the prediction and Planck limit, and the allowed region of $\theta_{23}$ is only around $\theta_{23} \simeq 52^\circ$.
For such a $\theta_{23}$, the sum of the neutrino masses $\sum_i m_i$ is so large that it implies the quasi-degenerate mass spectrum. 
In both the $\mathbf{D_1^R}$ and $\mathbf{D_2^R}$ cases, $\sum_i m_i$ is about $0.065$~eV and lighter than the limit.
But Eq.~\eqref{eq:cos-eq} has solutions for $\theta_{23} \gtrsim 44^\circ$ in the $\mathbf{D_1^R}$ case, and for $\theta_{23} \lesssim 46^\circ$ in the $\mathbf{D_2^R}$ case, as we see in Figs.~\ref{fig:mass-sum-s}~(\subref{fig:mass-D1R-NO}) and (\subref{fig:mass-D2R-NO}).
On the other hand, as we see in Figs.~\ref{fig:mass-sum-s}~(\subref{fig:mass-B3R-NO}), (\subref{fig:mass-B3R-IO}), (\subref{fig:mass-B4R-NO}) and (\subref{fig:mass-B4R-IO}), in these cases, the predicted values of  $\sum_i m_i$ conflict with the limit when we allow the parameters to be varied in $3\sigma$. Therefore these four cases are excluded.

From the above results, in the singlet cases, only three lepton flavor-dependent U(1) gauge symmetries, U(1)$_{L_\mu-L_\tau}~[\mathbf{C^R}]$, U(1)$_{B+L_e-3L_\mu-L_\tau}~[\mathbf{D_1^R}]$ and U(1)$_{B+L_e-L_\mu-3L_\tau}~[\mathbf{D_2^R}]$,  realize the NO without conflicting with the existing experiments and no symmetry does the IO. 
Furthermore, as also shown in Ref.~\cite{Asai:2018ocx}, the $\mathbf{C^R}$ case has strong tension with the Planck 2018 limit and will soon be tested in the future neutrino experiments.

On the other hand, in the doublet cases, only the U(1)$_{L_\mu-L_\tau}$ gauge symmetry realizes the two-zero texture structure and the possible mass orderings without conflicting with the various experiments relevant to $Z'$.
This case has been discussed in the previous work~\cite{Asai:2018ocx} and the results are shown there, but for completeness, we show them here.

\begin{figure}[htpb]
\begin{subfigure}[b]{0.5\linewidth}
\centering
\includegraphics[width=7cm]{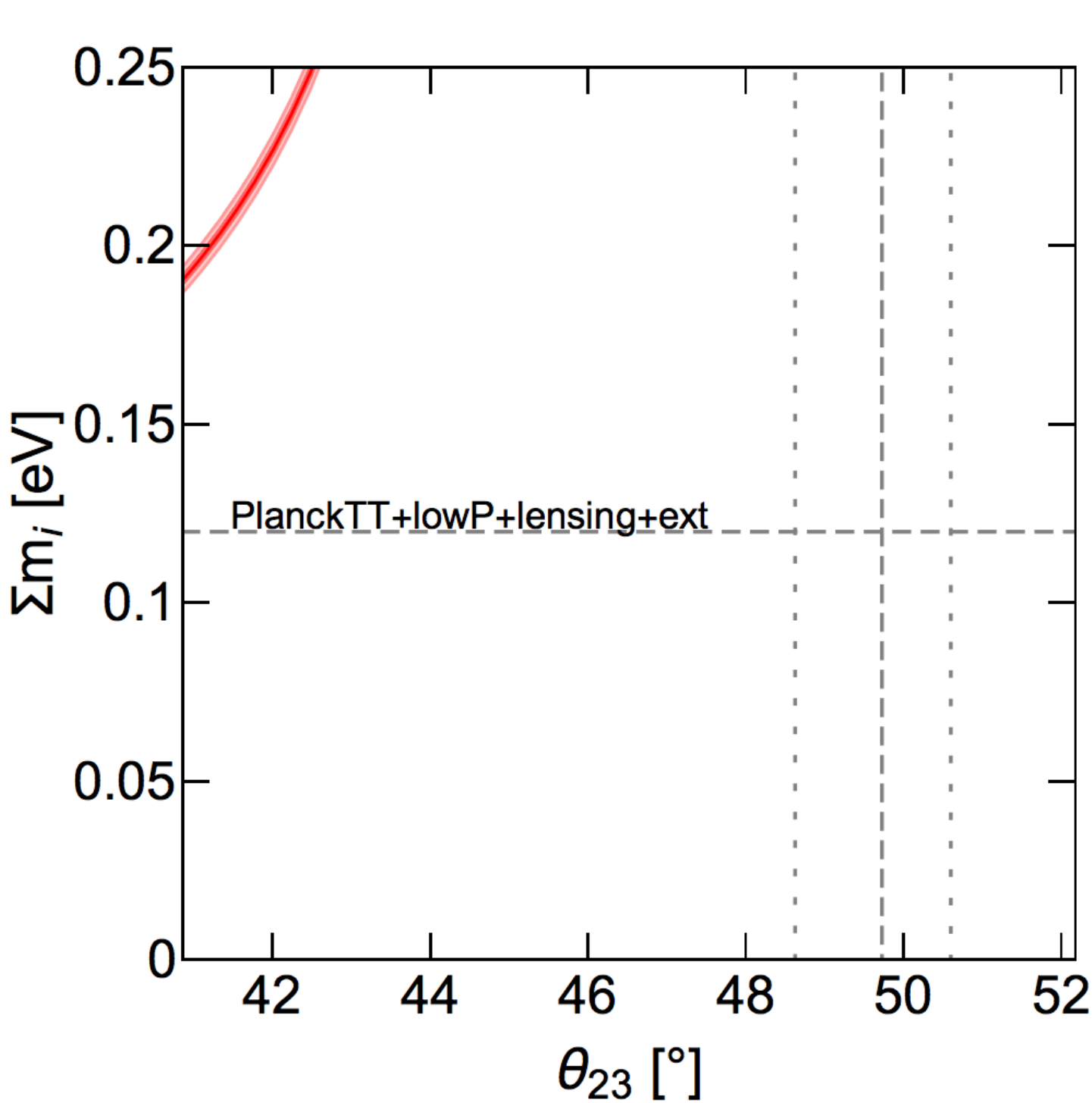}
\subcaption{$\mathbf{B_3^\nu}$ (NO)}
\label{fig:mass-B3nu-NO} 
\end{subfigure}
\begin{subfigure}[b]{0.5\linewidth}
\centering
\includegraphics[width=7cm]{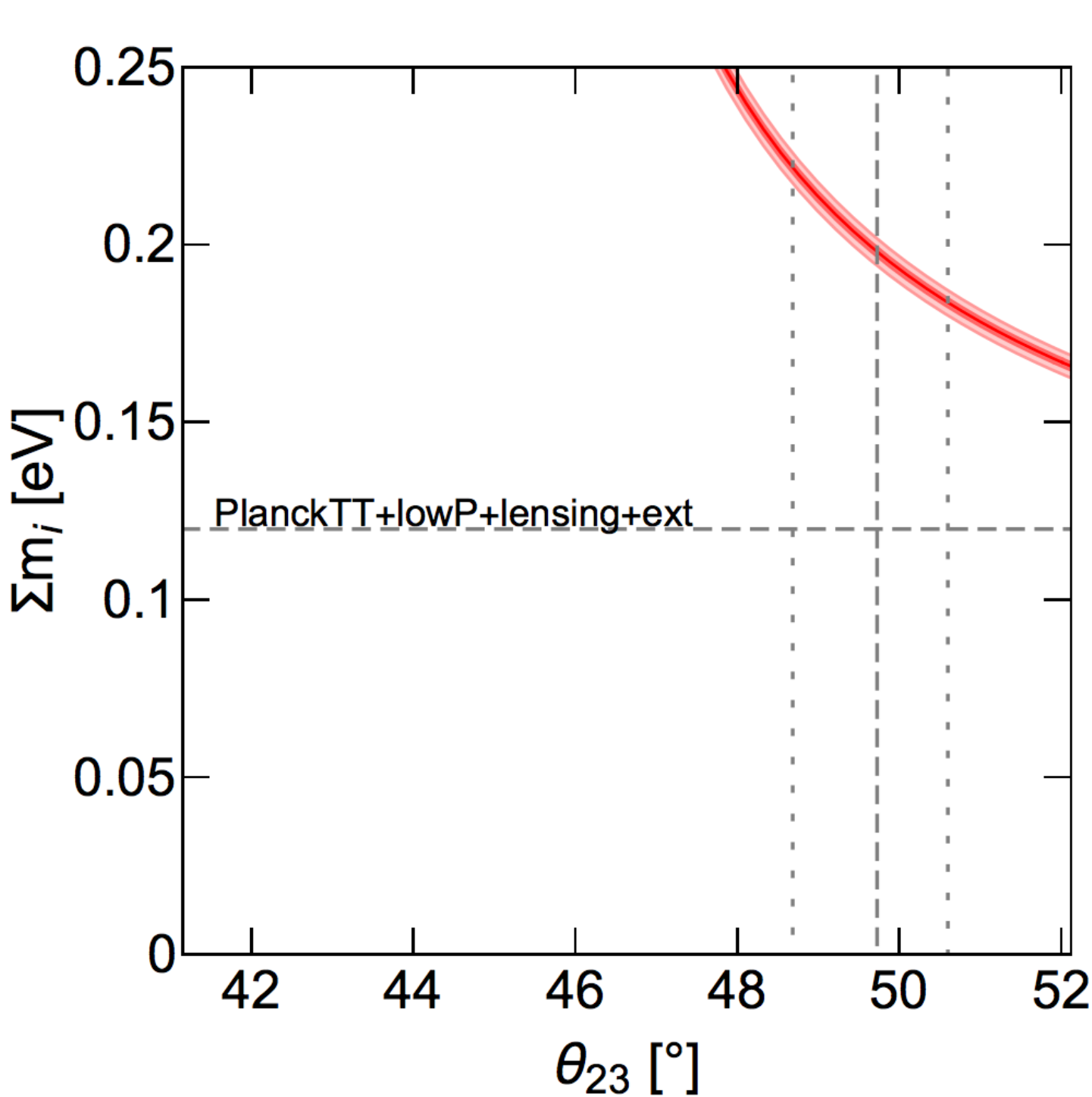}
\subcaption{$\mathbf{B_3^\nu}$ (IO)}
\label{fig:mass-B3nu-IO} 
\end{subfigure} \\
\begin{subfigure}[b]{0.5\linewidth}
\centering
\includegraphics[width=7cm]{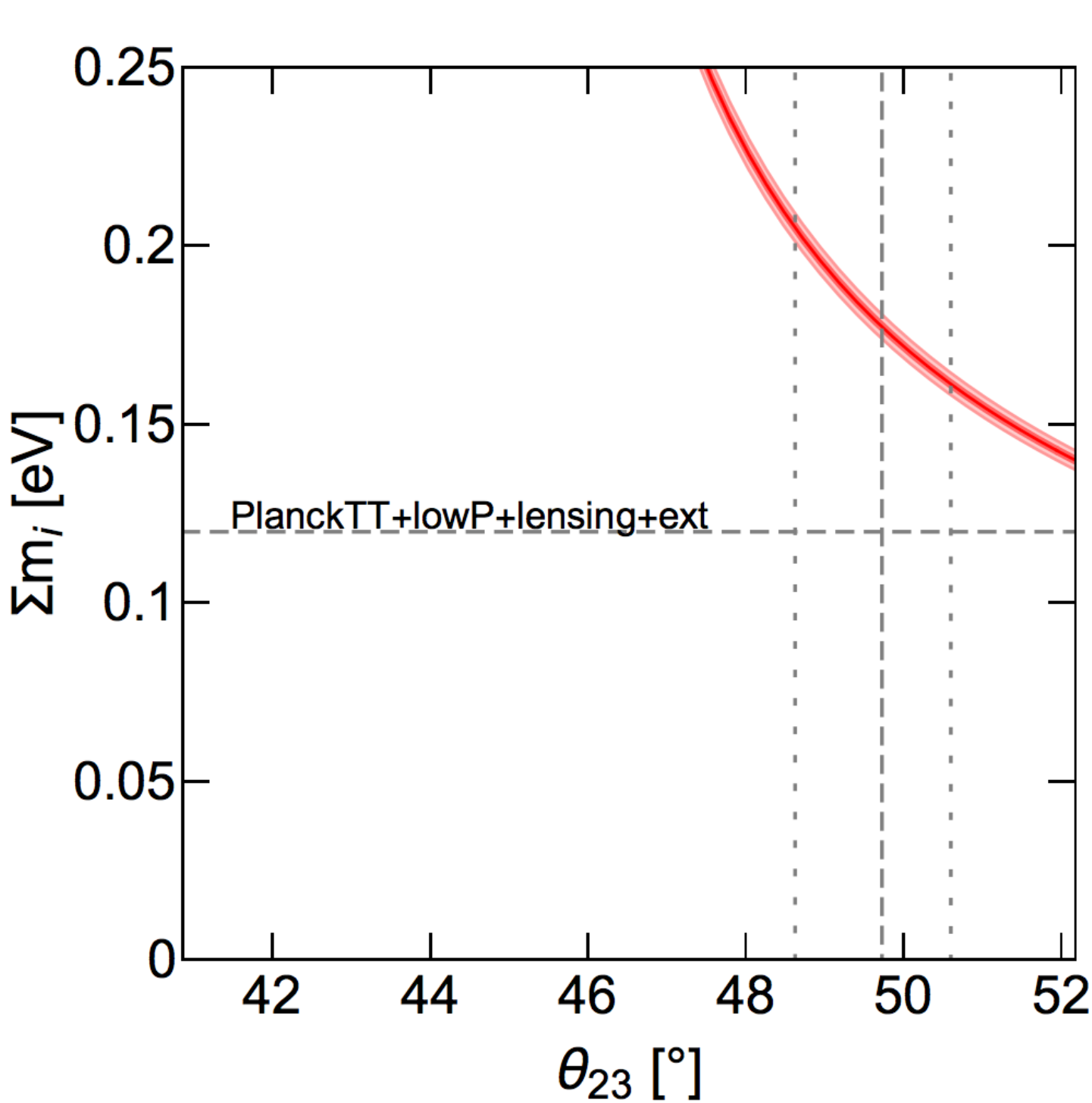}
\subcaption{$\mathbf{B_4^\nu}$ (NO)}
\label{fig:mass-B4nu-NO}
\end{subfigure}
\begin{subfigure}[b]{0.5\linewidth}
\includegraphics[width=7cm]{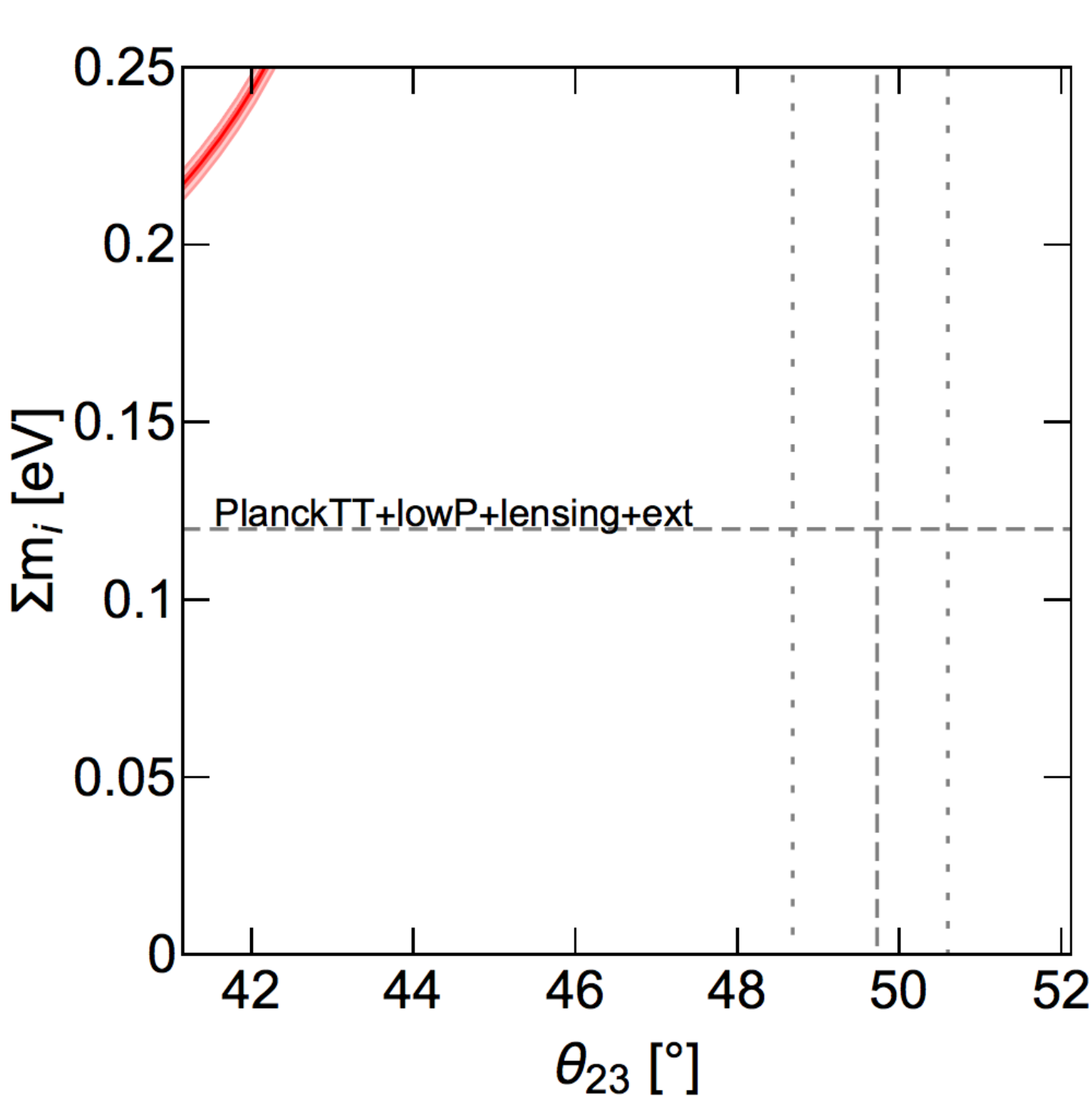}
\subcaption{$\mathbf{B_4^\nu}$ (IO)}
\label{fig:mass-B4nu-IO}
\end{subfigure}
\caption{
The predictions for the sum of the neutrino masses in the doublet models.  
The red line shows the prediction as functions of $\theta_{23}$ and the dark (light) red bands show the uncertainty coming from the $1\sigma$ ($3\sigma$) errors of $\Delta m^2_{3\ell}$. 
The vertical and horizontal lines are the same as Fig.~\ref{fig:mass-sum-s}.
}
\label{fig:mass-sum-d}
\end{figure}

In Fig.~\ref{fig:mass-sum-d}, which is the same as Fig.~2 in Ref.~\cite{Asai:2018ocx}, we plot the sum of the neutrino masses in the doublet models as functions of $\theta_{23}$ in the red line.
The dark (light) red bands show the uncertainty coming from the $1\sigma$ ($3\sigma$) errors of $\Delta m_{31}^2$.
The contribution to $\sum m_i$ from the errors of the other parameters are subdominant, and so we take them to be the best fit values.
Comparing Fig.~\ref{fig:mass-sum-d} with Fig.~\ref{fig:mass-sum-s}, we find that the predicted values in the $\mathbf{B_3^\nu}$ ($\mathbf{B_4^\nu}$) cases are the same as those in the $\mathbf{B_4^R}$ ($\mathbf{B_3^R}$).
This is because the inverse of the matrix which has the $\mathbf{B_3}$ structure has the $\mathbf{B_4}$ structure and vice versa.
\begin{align}
\renewcommand{\arraystretch}{0.8}
\begin{array}{ccc}
 \mathcal{M}_\nu &  & \mathcal{M}_\nu^{-1} \\ \hline
 \mathbf{B_3^\nu} : \left( \begin{array}{ccc} *&0&*\\0&0&*\\ *&*&* \end{array} \right) & \Longleftrightarrow &  \mathbf{B_4^R} : \left( \begin{array}{ccc} *&*&0\\ *&*&*\\0&*&0 \end{array} \right) \\ 
 \mathbf{B_4^\nu} : \left( \begin{array}{ccc} *&*&0\\ *&*&*\\0&*&0 \end{array} \right) & \Longleftrightarrow & \mathbf{B_3^R} : \left( \begin{array}{ccc} *&0&*\\0&0&*\\ *&*&* \end{array} \right) \\ \hline
\end{array} \nonumber 
\renewcommand{\arraystretch}{1}
\end{align}

In any case, as we have concluded in the previous work~\cite{Asai:2018ocx}, all the doublet cases result in too heavy neutrino masses and they are excluded by the Planck 2018 limit.
Therefore, we show the prediction of only the singlet cases hereafter.

\subsection{Dirac CP phase}
\label{subsec:result-dirac}
\begin{figure}[htpb]
\begin{subfigure}[b]{1\linewidth}
\centering
\includegraphics[width=7cm]{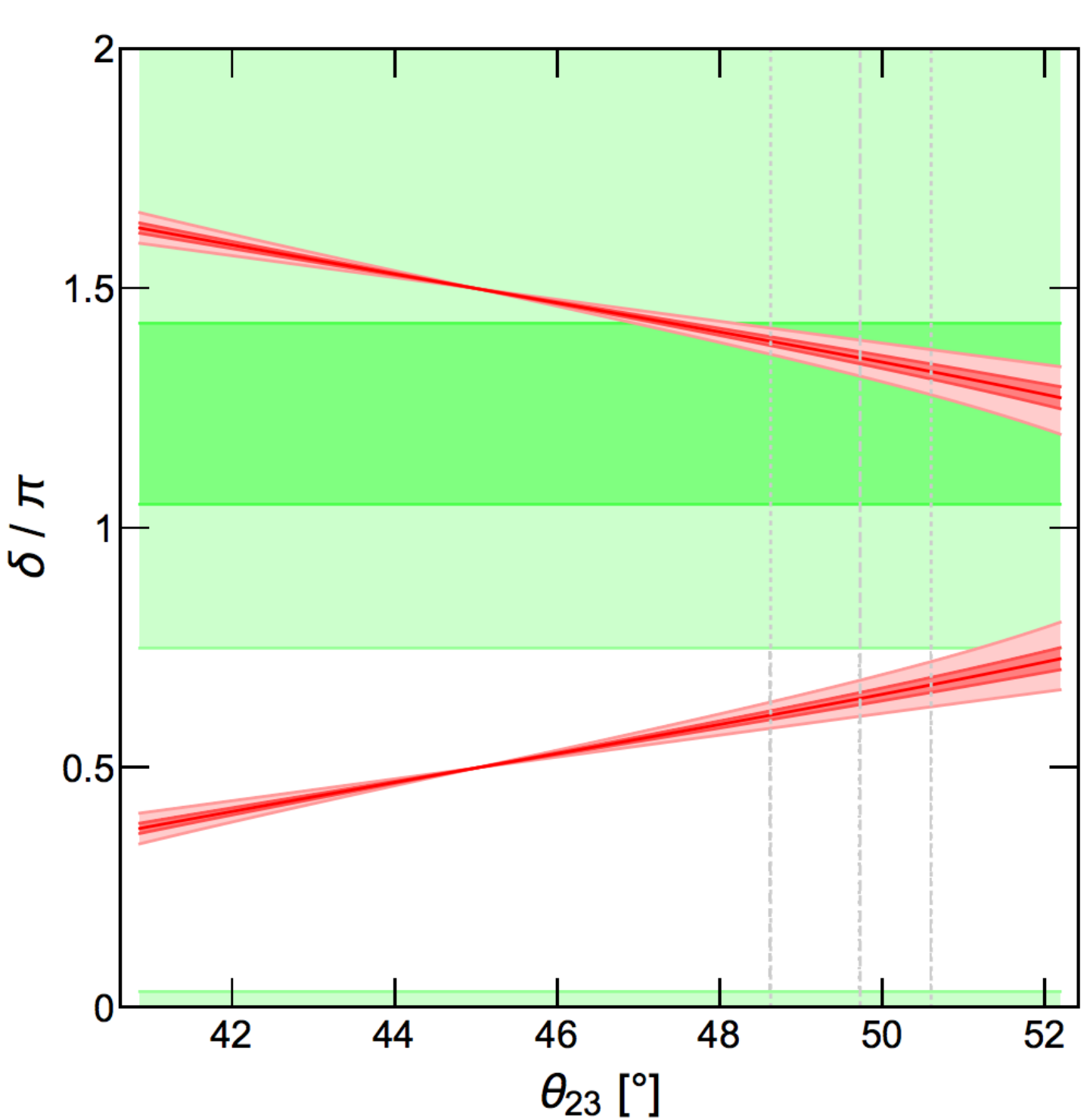}
\subcaption{$\mathbf{C^R}$ (NO)}
\label{fig:delta-CR-NO} 
\end{subfigure} \\ \\
\begin{subfigure}[b]{0.5\linewidth}
\centering
\includegraphics[width=7cm]{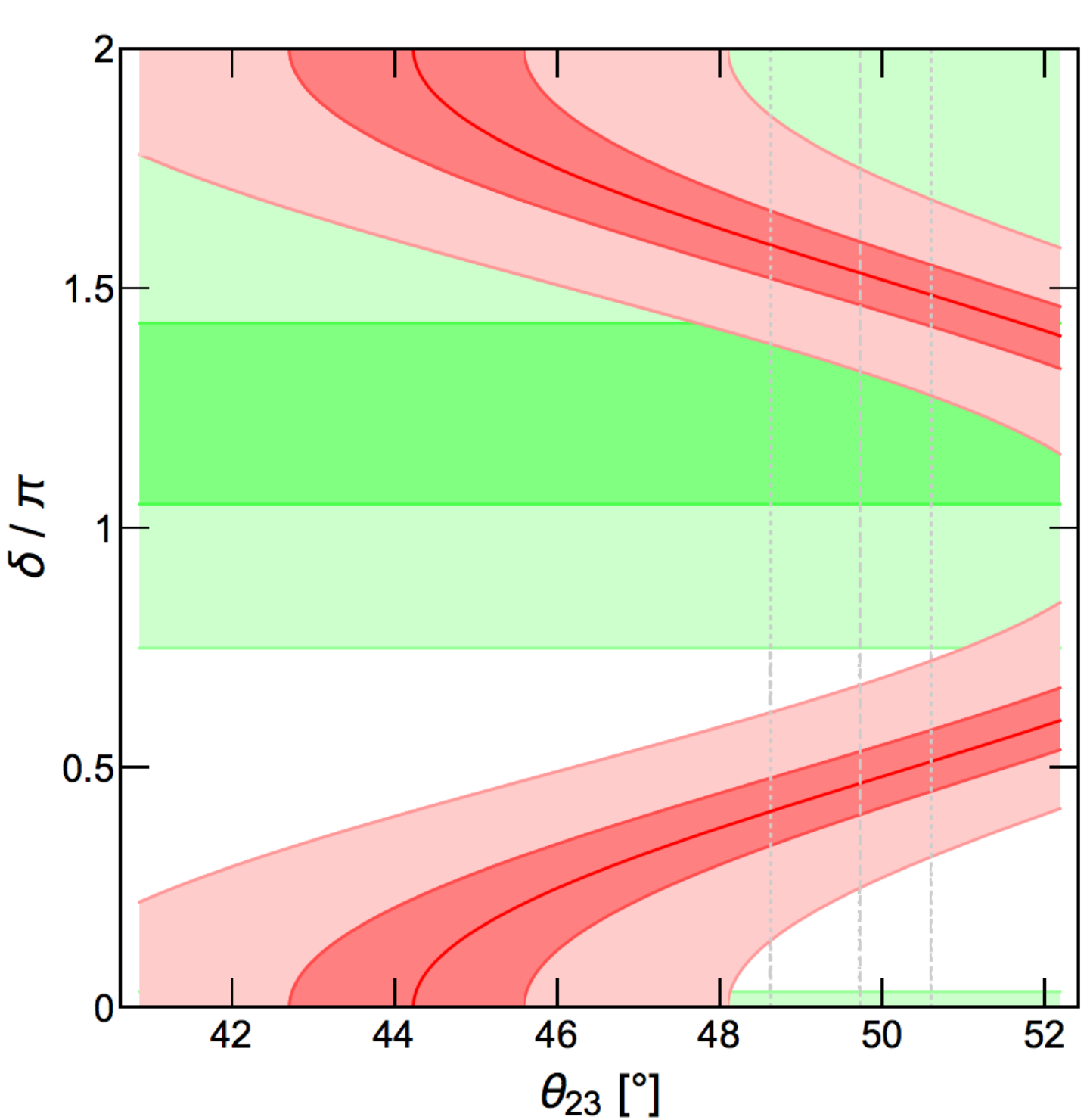}
\subcaption{$\mathbf{D_1^R}$ (NO)}
\label{fig:delta-D1R-NO}
\end{subfigure}
\begin{subfigure}[b]{0.5\linewidth}
\includegraphics[width=7cm]{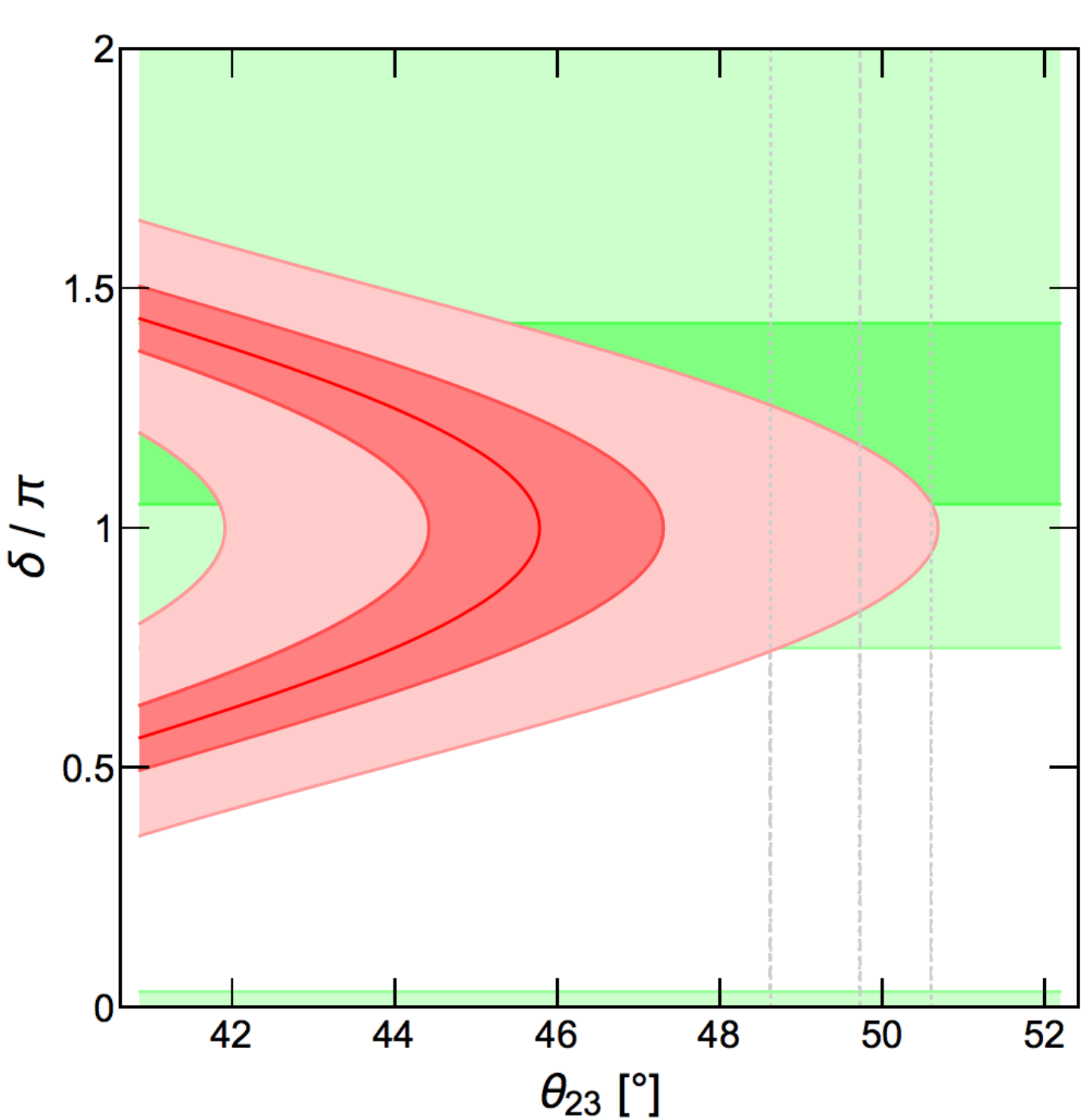}
\subcaption{$\mathbf{D_2^R}$ (NO)}
\label{fig:delta-D2R-NO}
\end{subfigure}
\caption{
The predictions for the Dirac CP phase in the singlet models.  
The red line shows the prediction as functions of $\theta_{23}$ and the dark (light) red bands show the uncertainty coming from the $1\sigma$ ($3\sigma$) errors of $\theta_{12}$ and the other parameters give subdominant contribution. 
The vertical gray lines are the same as Fig.~\ref{fig:mass-sum-s}. 
The horizontal dark (light) green bands show the $1\sigma$ ($3\sigma$) favored region of $\delta$.
}
\label{fig:delta-s}
\end{figure}

In Fig.~\ref{fig:delta-s}, we plot the Dirac CP phase as functions of $\theta_{23}$ in the red line. 
$\theta_{23}$ is varied in the $3\sigma$ region. 
Fig.~\ref{fig:delta-s}~(\subref{fig:delta-CR-NO}) is the same as Fig.~3 (a) in Ref.~\cite{Asai:2018ocx}.
The dark (light) red bands show the uncertainty coming from the $1\sigma$ ($3\sigma$) errors of $\theta_{12}$.
We notice that the dominant contribution to the uncertainty comes from the error in $\theta_{12}$.
On the other hand, the other parameters give subdominant contribution, and so we take them to be the best fit values. 
We also show the $1\sigma$ ($3\sigma$) favored region of $\delta$ in the dark (light) horizontal green bands.
As we see in Fig.~\ref{fig:delta-s}~(\subref{fig:delta-CR-NO}), in the $\mathbf{C^R}$ case, the upper band of the predicted values is in the experimentally favored region. For $\theta_{23} \simeq 52^\circ$, the prediction is $\delta \simeq 1.3 \pi$ and it is in the $1\sigma$ range of the favored region of $\delta$.
In the $\mathbf{D_1^R}$ and $\mathbf{D_2^R}$ cases, the predictions depend on $\theta_{12}$ as much as $\theta_{23}$ and the other parameters have considerable contributions.
Moreover, the quadratic equations of $\cos \delta$ have no solution in some regions of $\theta_{23}$ and $\mathbf{D_2^R}$ cannot realize the NO when we take all the neutrino oscillation parameters to be the best fit values.

\subsection{Effective Majorana neutrino mass}
\label{subsec:result-effmass}
\begin{figure}[htpb]
\centering
\includegraphics[width=7cm]{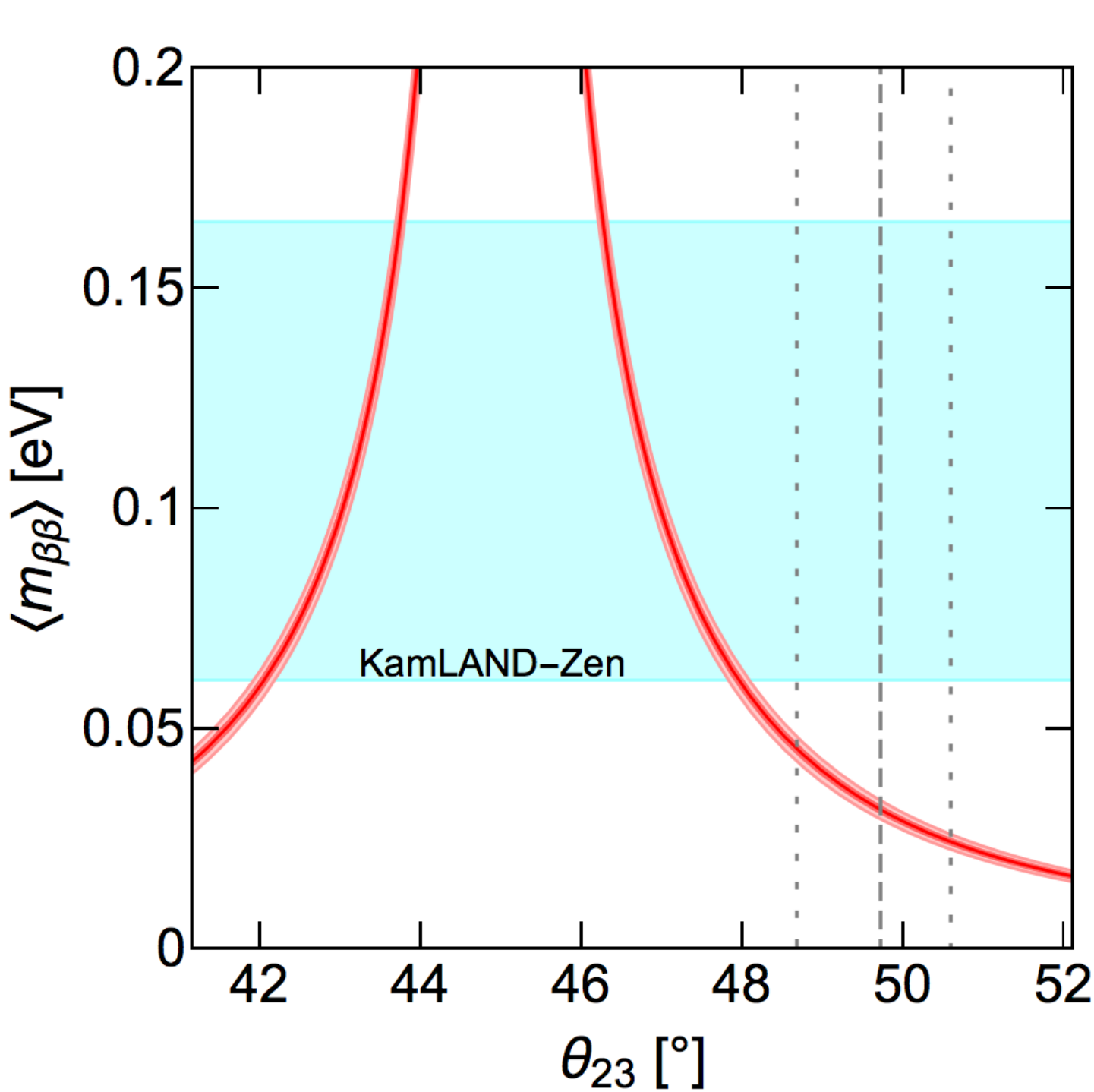}
\caption{
The predictions for the effective Majorana neutrino mass in the $\mathbf{C^R}$ case.  
The red line shows the prediction as functions of $\theta_{23}$ and the dark (light) red bands show the uncertainty coming from the $1\sigma$ ($3\sigma$) errors of the other parameters. 
The vertical gray lines are the same as Fig.~\ref{fig:mass-sum-s}. 
The light blue band shows the current bound on $\braket{m_{\beta \beta}}$ given by the KamLAND- Zen experiment, $\braket{m_{\beta \beta}}< 0.061\mathchar`-0.165$ eV~\cite{KamLAND-Zen:2016pfg}.
}
\label{fig:effmass-CR-NO}
\end{figure}

Lastly, we discuss neutrinoless double-beta decay ($0\nu \beta \beta$) and the possibility of testing the singlet models\footnote{See also Ref.~\cite{Crivellin:2015lwa}.}. 
The rate of neutrinoless double-beta decay is proportional to the square of the effective Majorana neutrino mass $\braket{m_{\beta \beta}}$, which is defined by
\begin{align}
\label{eq:effmass}
   \braket{m_{\beta\beta}} \equiv \left| \sum_i \left( U_{\text{PMNS}} \right)_{ei}^2 m_i \right| = \left| c_{12}^2 c_{13}^2 m_1 + s_{12}^2 c_{13}^2 e^{i\alpha_2} m_2 + s_{13}^2 e^{i(\alpha_3-2\delta)} m_3 \right|~.
\end{align}
As we note in the previous section, not only the neutrino mass $m_1$ and Dirac CP phase $\delta$, but also the Majorana CP phases $\alpha_{2,3}$ are uniquely determined as functions of the neutrino oscillation parameters $\theta_{12}$, $\theta_{23}$, $\theta_{13}$, $\Delta m_{21}^2$ and $\Delta m_{31}^2$.
Thus we can calculate the effective Majorana neutrino mass $\braket{m_{\beta \beta}}$ without ambiguity.
In Fig.~\ref{fig:effmass-CR-NO}, we show the predicted value of $\braket{m_{\beta \beta}}$, in the $\mathbf{C^R}$, as a function of $\theta_{23}$ in the red line.
This figure is also shown in Ref.~\cite{Asai:2018ocx}.
$\theta_{23}$ is varied in the $3\sigma$ region. 
The dark (light) red bands show the uncertainty coming from the $1\sigma$ ($3\sigma$) errors of the other parameters.
We also show in the light blue band the current bound on $\braket{m_{\beta \beta}}$ given by the KamLAND-Zen experiment, $\braket{m_{\beta \beta}}< 0.061\mathchar`-0.165$ eV~\cite{KamLAND-Zen:2016pfg}, where the uncertainty stems from the estimation of the nuclear matrix element for ${}^{136}$Xe.
As we see in Fig.~\ref{fig:effmass-CR-NO}, $\braket{m_{\beta \beta}}$ is predicted to be $\simeq 0.016$ eV for $\theta_{23} \simeq 52^\circ$, which is well below the present KamLAND-Zen limit.
Future experiments are expected to have sensitivities as low as $\mathcal{O}(0.01)$ eV~\cite{Agostini:2017jim}, and thus it is expected that the $\mathbf{C^R}$ case will be tested in the future.

On the other hand, the $\mathbf{D_1^R}$ and $\mathbf{D_2^R}$ cases predict that $\braket{m_{\beta \beta}}$ vanishes and it means that neutrinoless double-beta decay process never occurs.
This is because when the mass matrix of the neutrinos has the $\mathbf{D_1^R}$ ($\mathbf{D_2^R}$) structure, that matrix also has the $\mathbf{A_2^\nu}$ ($\mathbf{A_1^\nu}$) structure, respectively, and then ($e,e$) component of the neutrino mass matrix vanishes.
Thus the following relation is satisfied:
\begin{align}
   \braket{m_{\beta\beta}} &= \left| \left( U_{\text{PMNS}} \text{diag}(m_1,m_2,m_3) U_{\text{PMNS}}^T \right)_{ee} \right| \nonumber \\
   &= \left| \left( U_{\text{PMNS}}^* \text{diag}(m_1,m_2,m_3) U_{\text{PMNS}}^\dag \right)_{ee} \right| \nonumber \\
   &= \left| \left( \mathcal{M}_{\nu_L} \right)_{ee} \right| \nonumber \\
   &= 0~.
\end{align}
Therefore, the existing experimental limits on $\braket{m_{\beta\beta}}$ give no constraint to the $\mathbf{D_1^R}$ and $\mathbf{D_2^R}$ cases.
On the other hand, if neutrinoless double-beta decay processes are detected in future experiments, these cases are excluded.

\section{Implications for Leptogenesis}
\label{sec:LG}

The minimal gauged U(1)$_{Y'}$ models contain 3 right-handed neutrinos which couple to the SM leptons.
Therefore, if they are heavy enough, the observed baryon asymmetry of the Universe can be explained by the leptogenesis scenario~\cite{Fukugita:1986hr}, which is one of the most promising mechanisms.

In the previous section, we have shown that the neutrino masses, the Dirac CP phase, and Majorana CP phases are determined as functions of the neutrino oscillation parameters in the minimal gauged U(1)$_{Y'}$ models.
This smallness of the degree of freedom affects the parameters relevant to the baryon asymmetry generated through leptogenesis.

In this section, following the analyses of Ref.~\cite{Asai:2017ryy}, we firstly show the predicted signs of the asymmetry parameter $\epsilon$, which is one of the most important parameters in leptogenesis scenario, and then discuss whether the correct sign of the baryon asymmetry can be generated in the minimal gauged U(1)$_{Y'}$ models.
Next, we briefly discuss the possibility of generating enough amount of the baryon asymmetry in the thermal leptogenesis scenario.
For simplicity, we do not take into account flavor effects~\cite{Abada:2006fw,Nardi:2006fx,Abada:2006ea}.

In the minimal gauged U(1)$_{Y'}$ model, the absolute neutrino mass $m_1$, Dirac CP phase $\delta$ and Majorana CP phases $\alpha_{2,3}$ are not free parameters and determined as functions of the neutrino oscillation parameters $\theta_{12}, \theta_{23}, \theta_{13}, \Delta m_{21}^2, \Delta m_{31}^2$ and sign of $\sin \delta$.
Then, from the seesaw formula: $\mathcal{M}_{\nu_L} \simeq - \mathcal{M}_D \mathcal{M}_R^{-1} \mathcal{M}_D^T$, the Majorana mass matrix $\mathcal{M}_R$ is tightly constrained, and the free parameters in $\mathcal{M}_R$ are only the three Dirac Yukawa couplings of the neutrinos.
By diagonalizing the masses of the right-handed neutrinos, we can rewrite the Lagrangian \eqref{eq:lag} as follows:
\begin{align}
   \Delta \mathcal{L} = - \sum_{i=1}^3 \sum_{\alpha = e,\mu,\tau} \hat{\lambda}_{i\alpha} \hat{N}_i^c (L_\alpha \cdot H) - \frac{1}{2} \sum_{i=1}^3 M_i \hat{N}_i^c \hat{N}_i^c + h.c.~,
\end{align}
where $\hat{N}_i^c$ are the right-handed neutrinos in the basis where the Majorana mass matrix $\mathcal{M}_R$ is diagonal and $\hat{\lambda}_{i\alpha}$ are the Dirac Yukawa couplings of the neutrinos in that basis, and $M_i$ are the masses which are taken to be real and positive.
These values are written by
\begin{align}
   \mathcal{M}_R &= \Omega^* \text{diag}(M_1,M_2,M3) \Omega^\dag,\ \ \ \Omega^\dag \Omega = I~, \\
   \hat{N}_i^c &= \sum_\alpha \Omega_{\alpha i}^* N_\alpha^c~, \\
   \hat{\lambda}_{i\alpha} &= \Omega_{\alpha i} \lambda_\alpha \ \ \ (\text{not summed})~.
\end{align}
Here, for simplicity, we assume that the heavy neutrinos have mass hierarchy and the lightest right-handed neutrino $N_1$ is much lighter than the others.
In this case, the lepton asymmetry is generated by the decay of the lightest right-handed neutrino, and then the asymmetry parameter $\epsilon_1$ is given, at the leading order, by~\cite{Flanz:1994yx,Covi:1996wh,Buchmuller:1997yu}
\begin{align}
\label{eq:asy-para}
   \epsilon_1 &= \frac{1}{8\pi} \frac{1}{(\hat{\lambda} \hat{\lambda}^\dag)_{11}} \sum_{i=2,3} \text{Im} \left[ \left\{ (\hat{\lambda} \hat{\lambda}^\dag)_{1j} \right\}^2 \right] f \left( \frac{M_j^2}{M_1^2} \right)~, \\
   f(x) &= \sqrt{x} \left[ 1 - (x+1) \ln \left( 1 + \frac{1}{x} \right) - \frac{1}{x-1} \right]~.
\end{align}
As shown in Ref.~\cite{Asai:2017ryy}, this asymmetry parameter have a correlation with the sign of $\sin \delta$.
This is because that the Majorana CP phases flip: $\alpha_{2,3} \to -\alpha_{2,3}$ under sign flipping of the Dirac CP phase: $\delta \to -\delta$, and then the PMNS matrix transforms as $U_{\text{PMNS}} \to U_{\text{PMNS}}^*$.
On the other hand, the absolute neutrino mass $m_1$ depends on only $\cos \delta$, not $\sin \delta$, and so $m_1$ does not change under sign flipping of $\delta$.
Then $\mathcal{M}_{\nu_L}$, $\mathcal{M}_R$, $\Omega$, and $\hat{\lambda}$ also transform as $\mathcal{M}_{\nu_L} \to \mathcal{M}_{\nu_L}^*$, $\mathcal{M}_R \to \mathcal{M}_R^*$, $\Omega \to \Omega^*$, and  $\hat{\lambda} \to \hat{\lambda}^*$.
Therefore, from Eq.~\eqref{eq:asy-para}, the sign of the asymmetry parameter $\epsilon_1$ is flipped: $\epsilon_1 \to -\epsilon_1$, and it means that there is a one-to-one correspondence between the sign of the Dirac CP phase and that of the baryon asymmetry in the Universe. In the following discussion, we take the Dirac CP phase to be $\delta > \pi$, which is favored by the experiments.

When we focus on only the sign of the baryon asymmetry, only the ratio between the Yukawa couplings $\lambda_e$, $\lambda_\mu$, $\lambda_\tau$ are important.
 Now we parametrize the Yukawa couplings as follows:
 \begin{align}
    (\lambda_e, \lambda_\mu, \lambda_\tau)^T = \lambda ( \cos \theta, \sin \theta \cos \phi, \sin \theta \sin \phi )^T \equiv \lambda \mathbf{n}^T~.
\end{align}
For this parametrization, the asymmetry parameter in Eq.~\eqref{eq:asy-para} is written by
\begin{align}
\label{eq:asy-para2}
   \epsilon_1 &= \frac{1}{8\pi} \frac{\lambda^2}{(\hat{n} \hat{n}^\dag)_{11}} \sum_{i=2,3} \text{Im} \left[ \left\{ (\hat{n} \hat{n}^\dag)_{1j} \right\}^2 \right] f \left( \frac{M_j^2}{M_1^2} \right)~, \\
   \hat{n}_{i\alpha} &= \Omega_{\alpha i} n_\alpha \ \ \ (\text{not summed})~,
\end{align}
From Eq.~\eqref{eq:asy-para2}, the sign of $\epsilon_1$ is independent of the scale of the Yukawa couplings $\lambda$, and it is determined by the ratios $\theta$, $\phi$.
Note that the sphaleron process predicts $n_B/n_L < 0$~\cite{Harvey:1990qw}, and then $\epsilon_1 < 0$ leads to the observed baryon asymmetry of the Universe.

\begin{figure}[htpb]
\begin{subfigure}[b]{1\linewidth}
\centering
\includegraphics[width=7cm]{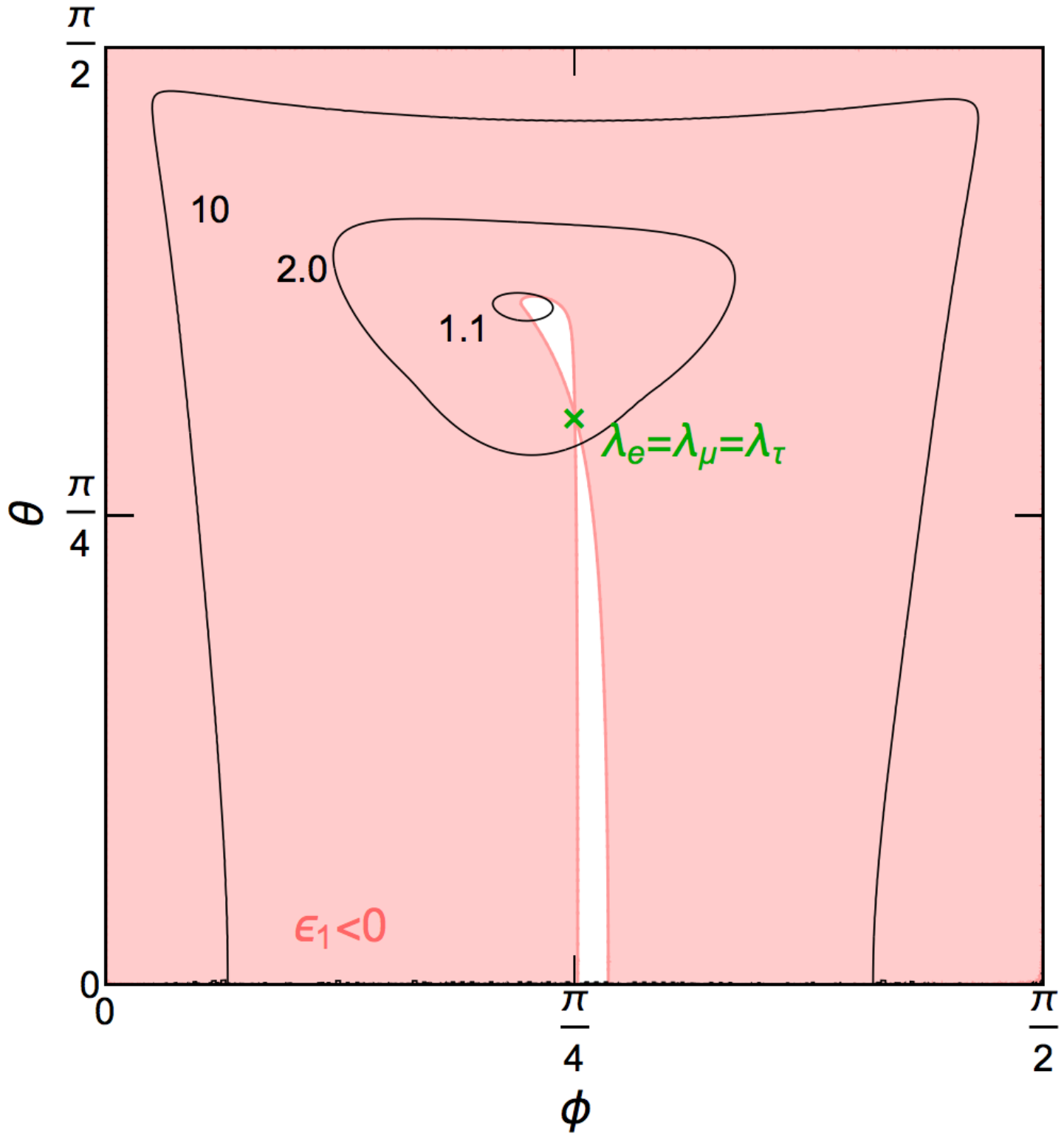}
\subcaption{$\mathbf{C^R}$ (NO)}
\label{fig:sign-CR-NO} 
\end{subfigure} \\
\begin{subfigure}[b]{0.5\linewidth}
\centering
\includegraphics[width=7cm]{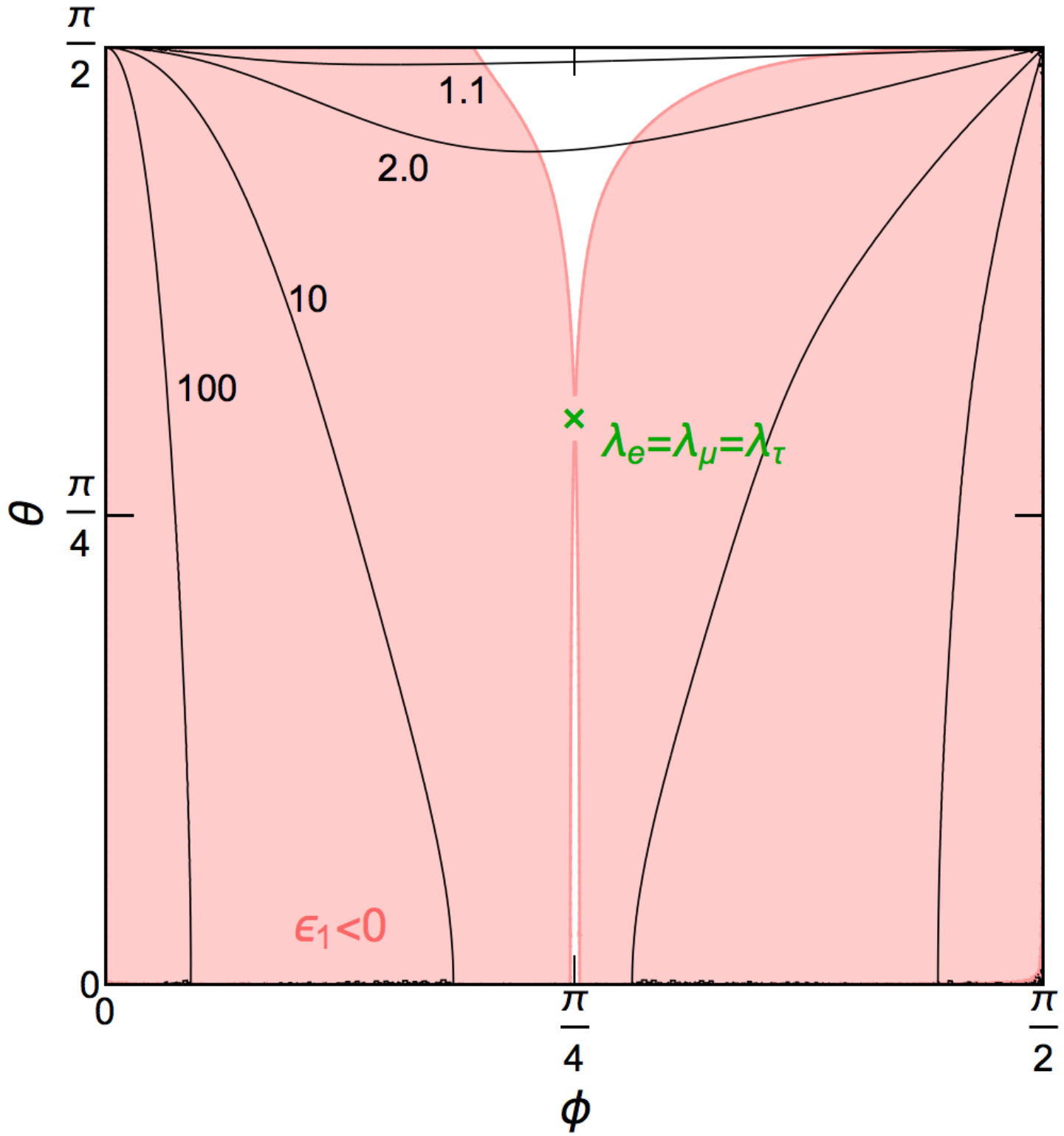}
\subcaption{$\mathbf{D_1^R}$ (NO)}
\label{fig:sign-D1R-NO}
\end{subfigure}
\begin{subfigure}[b]{0.5\linewidth}
\includegraphics[width=7cm]{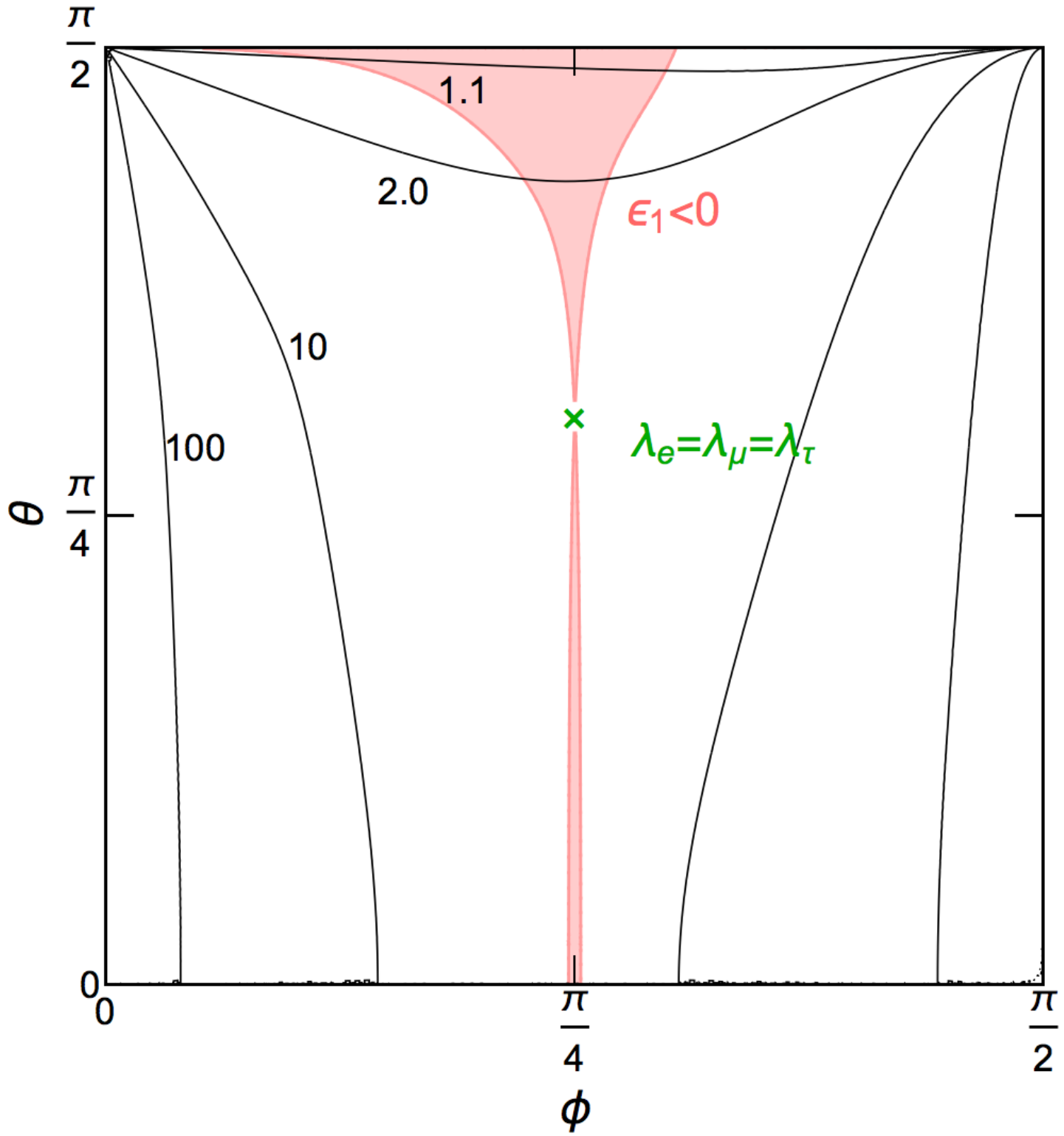}
\subcaption{$\mathbf{D_2^R}$ (NO)}
\label{fig:sign-D2R-NO}
\end{subfigure}
\caption{
The predictions for the sign of the baryon asymmetry in the singlet models. 
The red shaded areas show the parameter region of the $(\theta,\phi)$ plane where the asymmetry parameter $\epsilon_1$ is negative: $\epsilon_1 < 0$.
Moreover, we also show the contours of the right-handed neutrino mass ratio $M_2/M_1$. 
In Figs.~(\subref{fig:sign-CR-NO}), (\subref{fig:sign-D1R-NO}) and (\subref{fig:sign-D2R-NO}), we use $\theta_{23} = 52^\circ, 49.7^\circ \text{(best fit value)}, 43^\circ$, respectively, and take the other parameters to be the best fit values.
}
\label{fig:sign-epsilon}
\end{figure}

In Fig.~\ref{fig:sign-epsilon}, we show the parameter region of the $(\theta,\phi)$ plane where the asymmetry parameter $\epsilon_1$ is negative.
In Figs.~\ref{fig:sign-epsilon}~(\subref{fig:sign-CR-NO}), (\subref{fig:sign-D1R-NO}) and (\subref{fig:sign-D2R-NO}), we use $\theta_{23} = 52^\circ, 49.7^\circ \text{(best fit value)}, 43^\circ$, respectively, and take the other parameters to be the best fit values.
The red shaded areas show the parameter region of the $(\theta,\phi)$ plane where the asymmetry parameter $\epsilon_1$ is negative: $\epsilon_1 < 0$.
Moreover, we also show the contours of the right-handed neutrino mass ratio $M_2/M_1$.
These figures show that the $\mathbf{C^R}$ and $\mathbf{D_1^R}$ cases predict $\epsilon_1<0$ in the large regions of the $(\theta, \phi)$ plane.
On the other hand, $\mathbf{D_2^R}$ case realizes $\epsilon_1 < 0$ only on the thin red line: $\phi \simeq \pi/4$ \& $\theta \lesssim \pi/4$ ($\lambda_\mu \simeq \lambda_\tau < \lambda_e$) and in the red triangle: $\theta > \pi/4$.

For the estimation of the final baryon asymmetry generated by the leptogenesis scenario, the production mechanism of the right-handed neutrinos and the efficiency factor $\kappa_f$ are important. 
In the case of thermal leptogenesis, there are weak and strong washout regimes.
The $\mathbf{C^R}$ case predicts $m_1 \gtrsim 0.03$~eV, and the $\mathbf{D_1^R}$ and $\mathbf{D_2^R}$ cases do $m_1 \gtrsim 0.004$~eV, and then they are classified in the strong washout regime\footnote{See , e.g., \cite{Buchmuller:2004nz}}.
In that regime, the efficiency factor $\kappa_f$ is given by~\cite{Buchmuller:2004nz}
\begin{align}
\label{eq:efficiency}
   \kappa \simeq 0.02 \left( \frac{0.01~\text{eV}}{\tilde{m}_1} \right)^{1.1},
\end{align}
where $\tilde{m}_1 \equiv (\mathcal{M}^\dag \mathcal{M}_D)_{11}/M_1$ is the effective neutrino mass.
In general, the relation $\tilde{m}_1 \geq m_1$ holds~\cite{Fujii:2002jw}, and then $\kappa_f$ are roughly given as follows;
\begin{align}
   \kappa_f \lesssim \left\{ \begin{array}{ll} 0.02 \times \left( \frac{0.01~\text{eV}}{0.03~\text{eV}} \right)^{1.1} \simeq 0.006 & (\mathbf{C^R}) \\ 0.02 \times \left( \frac{0.01~\text{eV}}{0.004~\text{eV}} \right)^{1.1} \simeq 0.055 & (\mathbf{D_1^R, D_2^R}) \end{array} \right. .
\end{align}
The generated baryon asymmetry is given by~\cite{Buchmuller:2004nz}
\begin{align}
   Y_{\Delta B} = - \frac{28}{79} \frac{\kappa_f}{g_*} \epsilon_1 \simeq -3.3 \times 10^{-3} \kappa_f \epsilon_1,
\end{align}
 where $g_*$ is the degree of freedom of relativistic fields.
 Therefore, the asymmetry parameter needs to be $\epsilon_1 \sim \mathcal{O}(10^{-5} \mathchar`- 10^{-6})$ to explain the observed baryon asymmetry $Y_{\Delta B} \sim 8.7\times 10^{-11}$~\cite{Aghanim:2018eyx}.
 
\begin{figure}[htpb]
\begin{subfigure}[b]{1\linewidth}
\centering
\includegraphics[width=7cm]{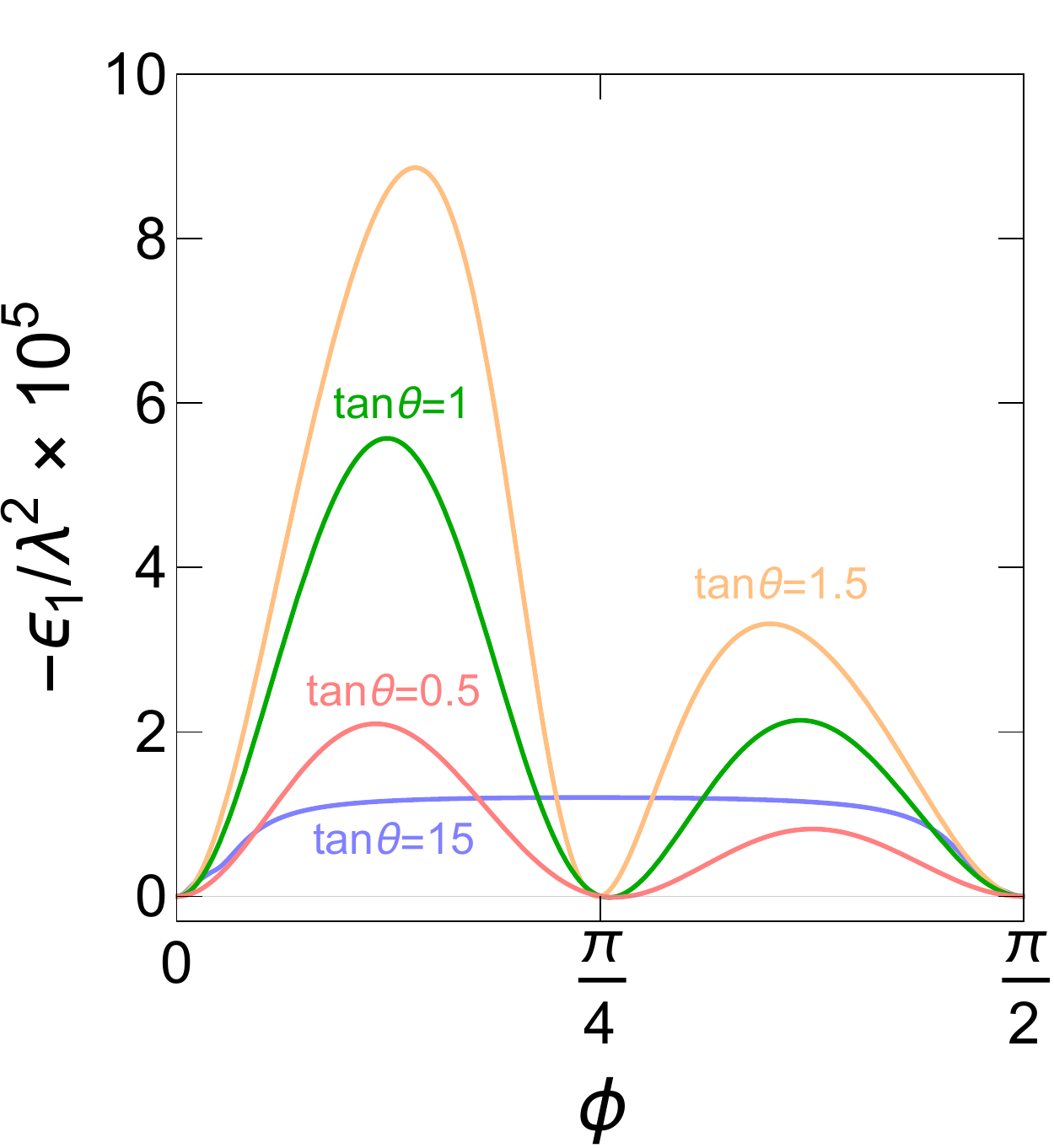}
\subcaption{$\mathbf{C^R}$ (NO)}
\label{fig:size-CR-NO} 
\end{subfigure} \\
\begin{subfigure}[b]{0.5\linewidth}
\centering
\includegraphics[width=7cm]{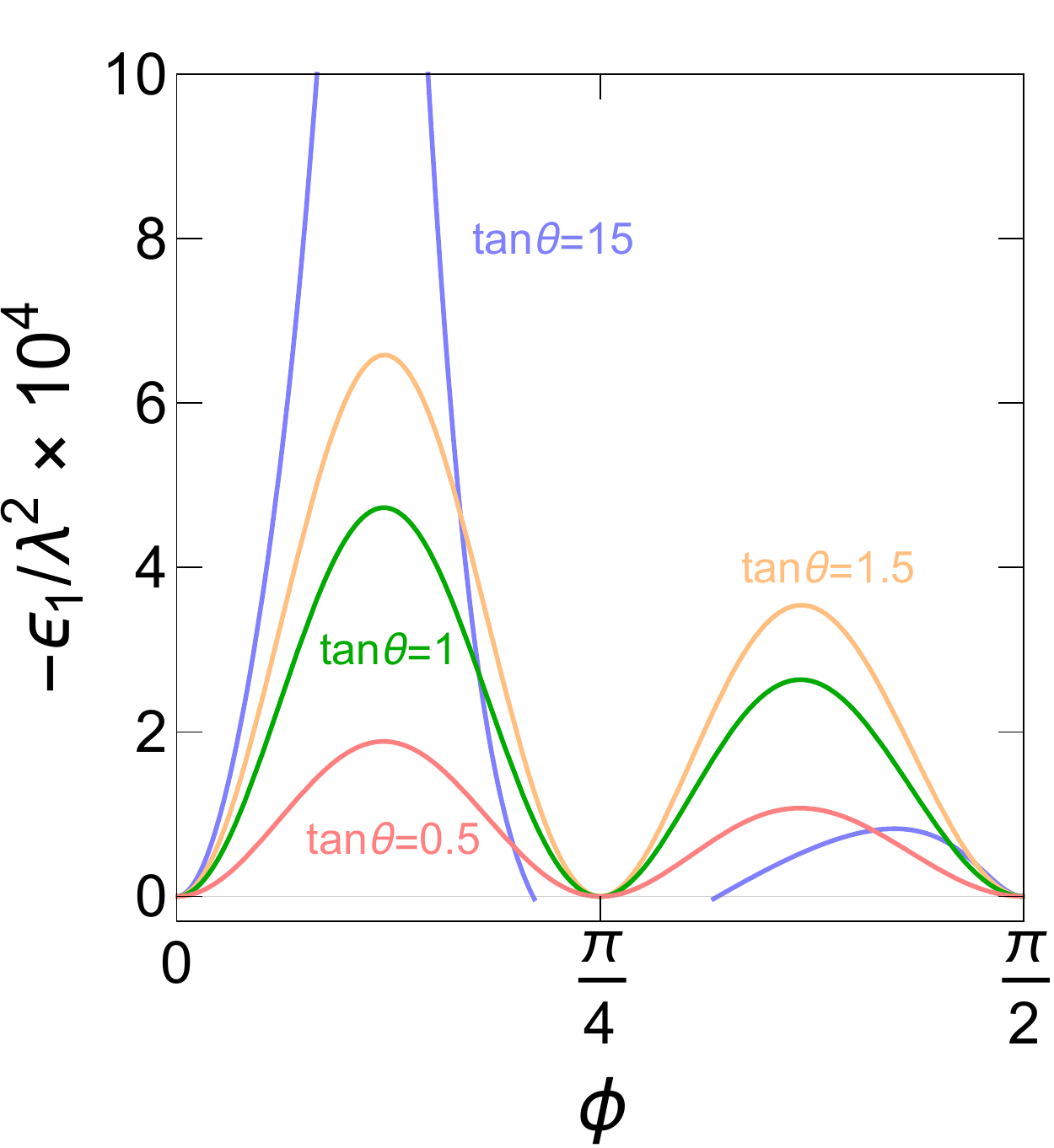}
\subcaption{$\mathbf{D_1^R}$ (NO)}
\label{fig:size-D1R-NO}
\end{subfigure}
\begin{subfigure}[b]{0.5\linewidth}
\includegraphics[width=7cm]{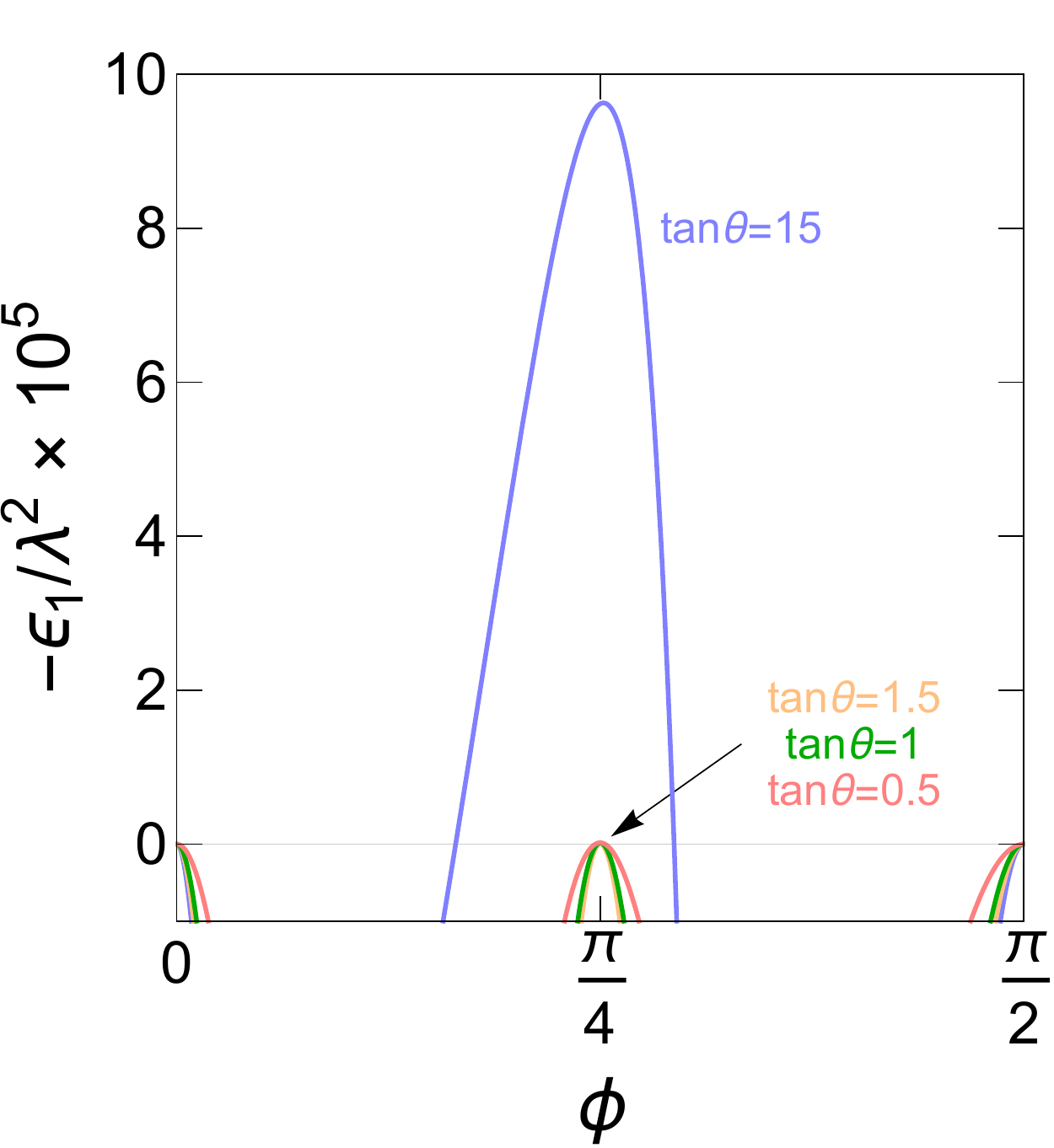}
\subcaption{$\mathbf{D_2^R}$ (NO)}
\label{fig:size-D2R-NO}
\end{subfigure}
\caption{
The predictions for the sizes of the asymmetry parameters $-\epsilon_1/\lambda^2$ as functions of the ratios of the Yukawa couplings $\theta, \phi$ in the singlet models. 
In Figs.~(\subref{fig:size-CR-NO}), (\subref{fig:size-D1R-NO}) and (\subref{fig:size-D2R-NO}), we use $\theta_{23} = 52^\circ, 49.7^\circ \text{(best fit value)}, 43^\circ$, respectively, and take the other parameters to be the best fit values.
}
\label{fig:size-epsilon}
\end{figure}

 In Figs.~\ref{fig:size-epsilon}, we show the sizes of the asymmetry parameters $-\epsilon_1/\lambda^2$ as functions of $\phi$. 
As you see in Eq.~\eqref{eq:asy-para2}, for fixed $\theta$, the asymmetry parameter $\epsilon_1$ scales as $\epsilon_1 \propto \lambda^2$ in this parametrization, and so we normalize $\epsilon_1$ by $\lambda^2$.
The neutrino oscillation parameters except for $\theta_{23}$ are taken to be the best fit values, and $\theta_{23}$ are taken to be $\theta_{23} = 52^\circ, 49.7^\circ \text{(best fit value)}, 43^\circ$, respectively. 
The sign of the Dirac CP phase is taken to be negative ($\delta > \pi$).
In this setup, these figures show that the $\mathbf{C^R}$ and $\mathbf{D_1^R}$ cases predict sizable values $|\epsilon_1| > 10^{-5}$.
On the other hand, it is difficult for the $\mathbf{D_2^R}$ case to give a sizable negative value, and only for $2 \lesssim \tan \theta \lesssim 200$, that is, in the red triangle area in Fig.~\ref{fig:sign-epsilon}~(\subref{fig:sign-D2R-NO}), the predicted value of $-\epsilon_1/\lambda^2$ is greater than $10^{-6}$.
In a part of this area, as we see in Fig.~\ref{fig:sign-epsilon}~(\subref{fig:sign-D2R-NO}), the right-handed neutrino masses $M_1, M_2$ are degenerate, and so we have to consider the resonant leptogenesis~\cite{Pilaftsis:2003gt,Blanchet:2006dq,Engelhard:2006yg,Dev:2017wwc} for more correct discussion.

\section{Conclusions}
\label{sec:conclusion}

In this work, we have extended U(1)$_{L_\alpha-L_\beta}$ gauge symmetries in Refs.~\cite{Asai:2017ryy,Asai:2018ocx} to more general lepton flavor-dependent U(1)$_{Y'}$, namely linear combinations of U(1)$_{L_e-L_\mu}$, U(1)$_{L_\mu-L_\tau}$ and U(1)$_{B-L}$, and have studied the two-zero minor and texture structures in the minimal gauged U(1)$_{Y'}$ models. 
In these models, because of the gauge symmetries, the structures of the Dirac and Majorana mass matrices are tightly restricted.
These restrictions connect the low-energy neutrino parameters each other, and four of them are given as functions of the rest of the parameters, namely the neutrino mixing angles and squared mass differences.
Using these relations, we obtained the prediction for the Dirac CP phase, sum of the neutrino masses, and effective Majorana neutrino mass. 
Furthermore, among the minimal U(1)$_{Y'}$ gauged models which realize the TZM or TZT, we found that only three models with one U(1)$_{Y'}$-breaking singlet scalar for NO are not in conflict with the experimental constraints.
We summarize the results in Tab.~\ref{tab:summary}.
``N'' and ``I'' represent the NO and IO, respectively. 
``$\checkmark$'' and ``$\times$'' represent whether models are consistent with the experimental constraints or not. 
``$\triangle$'' means that, in the column of mixing, the models predict the NO or the IO if $\theta_{23}$ is taken to be in the range of $3\sigma$, and in that of mass, there is a strong tension between the prediction and the Planck limit.
Moreover, in that of leptogenesis, it means that the limited parameter regions realize the correct sign and enough size of the baryon asymmetry. 
``$\square$'' means that the models predict that the $0\nu \beta \beta$ process never occurs.
\begin{table}[t]
\centering
\caption{
The consistency between the models with the TZM structure and the experimental constraints, namely the neutrino mixing angles, sum of the neutrino masses and effective Majorana neutrino mass. 
We also show the possibility to generate the correct sign of the baryon asymmetry. 
``N'' and ``I'' represent the NO and IO, respectively. 
``$\checkmark$'' and ``$\times$'' represent whether models are consistent with the experimental constraints or not. 
``$\triangle$'' means that, in the column of mixing, the models predict the NO or the IO if $\theta_{23}$ is taken to be in the range of $3\sigma$, and in that of mass, there is a strong tension between the prediction and the Planck limit.
Moreover, in that of leptogenesis, it means that the limited parameter regions realize the correct sign and enough size of the baryon asymmetry. 
``$\square$'' means that the models predict that the $0\nu \beta \beta$ process never occurs.
}
\label{tab:summary}
\vspace{12pt}
\begin{tabular}{llllllllllllllllll} \hline 
& \multicolumn{17}{l}{Structural index} \\ \cline{2-18}
  & $\mathbf{A_1^R}$ & $\mathbf{A_2^R}$ & \multicolumn{2}{c}{$\mathbf{B_3^R}$}  & \multicolumn{2}{c}{$\mathbf{B_4^R}$} & \multicolumn{2}{c}{$\mathbf{C^R}$} & \multicolumn{2}{c}{$\mathbf{D_1^R}$} & \multicolumn{2}{c}{$\mathbf{D_2^R}$} & $\mathbf{E_1^R}$ & $\mathbf{E_2^R}$ & $\mathbf{F_1^R}$ & $\mathbf{F_2^R}$ & $\mathbf{F_3^R}$ \\
 & N$\cdot$I & N$\cdot$I & N & I & N & I & N & I & N & I & N & I & N$\cdot$I &N$\cdot$I & N$\cdot$I & N$\cdot$I & N$\cdot$I \\ \hline
 Mixing & $\times$ & $\times$ & $\checkmark$ & $\triangle$ & $\triangle$ & $\checkmark$ & $\checkmark$ & $\times$ & $\checkmark$ & $\times$ & $\checkmark$ & $\times$ & $\times$ & $\times$ & $\times$ & $\times$ & $\times$ \\
 Mass & $-$ & $-$ & $\times$ & $\times$ & $\times$ & $\times$ & $\triangle$ & $-$ & $\checkmark$ & $-$ & $\checkmark$ & $-$ & $-$ & $-$ & $-$ & $-$ & $-$ \\ 
 $0\nu \beta \beta$ & $-$ & $-$ & $-$ & $-$ & $-$ & $-$ & $\checkmark$ & $-$ & $\square$ & $-$ & $\square$ & $-$ & $-$& $-$ & $-$ & $-$ & $-$ \\ 
 Leptogenesis & $-$ & $-$ & $-$ & $-$ & $-$ & $-$ & $\checkmark$ & $-$ & $\checkmark$ & $-$ & $\triangle$ & $-$ & $-$ & $-$ & $-$ & $-$ & $-$ \\ \hline 
\end{tabular}
\end{table}

Among them, U(1)$_{L_\mu-L_\tau} [\mathbf{C^R}]$ case predicts heavy neutrino masses and has a strong tension with the Planck 2018 limit. Therefore, this case will soon be tested by the future experiments of $\sum_i m_i$ and $\theta_{23}$.
On the other hand, U(1)$_{B+L_e-3L_\mu-L_\tau} [\mathbf{D_1^R}]$ and U(1)$_{B+L_e-L_\mu-3L_\tau} [\mathbf{D_2^R}]$ cases predict lighter $\sum_i m_i$ than the Planck limit and no neutrinoless double beta decay.
Then if neutrinoless double beta decay is observed in the future, these cases will be excluded completely. 

We have also discussed the implications of the minimal gauged U(1)$_{Y'}$ models for the leptogenesis scenario.
Because of the two-zero minor conditions, the neutrino Majorana mass matrix in the minimal models is severely restricted and have only three free parameters.
We found that the $\mathbf{C^R}$ and $\mathbf{D_1^R}$ cases generate the correct sign of the baryon asymmetry through the leptogenesis scenario in the large parameter spaces, and predict enough size of the asymmetry parameters in the thermal leptogenesis scenario.
On the other hand, the $\mathbf{D_2^R}$ case generates the correct sign of the baryon asymmetry only in the limited regions of the parameter space, and the observed value of the baryon asymmetry can be generated in this case when the right-handed neutrino masses are degenerate.

\section*{Acknowledgments}
We thank Koichi Hamaguchi and Natsumi Nagata for valuable discussions and suggestions.
This work was supported by JSPS KAKENHI Grant Number 19J13812.

\section*{Data Availability Statement}
This manuscript has no associated data or the data will not be deposited.
[Authors’ comment: The values of the neutrino oscillation parameters used during this analysis are shown in Tab.~\ref{tab:input}, and they can be obtained from Ref.~\cite{nufit, Esteban:2018azc}]

\newpage
\section*{Appendix}
\appendix

\section{Miscellaneous formulae}
\label{sec:misc}
\renewcommand{\theequation}{a.\arabic{equation}}
\setcounter{equation}{0}

Here we give formulae that are useful for the study of the neutrino mass structures in three minimal gauged U(1)$_{Y'}$ models which do not conflict with the experimental results, such as the neutrino oscillation experiments and the Planck 2018. 
The formulae in the $\mathbf{C^R}$ case are also shown in Ref.~\cite{Asai:2017ryy}.

\subsection{$R_2$ and $R_3$}
\label{app:r2r3}

Here, we show the list of the functions relevant to the ratios $m_{2,3}/m_1$, namely $R_{2,3}(\delta)$ in Tab.~\ref{tab:R23}.

\begin{table}[h]
\centering
\caption{
The list of the functions $R_{2,3}$.
}
\label{tab:R23}
\vspace{5pt}
\begin{tabular}{|c|c|c|} \hline \hline
structure index & $R_i$ &  \\ \hline 
\multirow{2}{*}{$\mathbf{C^R}$} & $R_2$ & $- \frac{2 \sin^2 \theta_{12} \cos 2\theta_{23} + \sin 2\theta_{12} \sin 2\theta_{23} \sin \theta_{13} e^{i\delta}}{2 \cos^2 \theta_{12} \cos 2\theta_{23} - \sin 2\theta_{12} \sin 2\theta_{23} \sin \theta_{13} e^{i\delta}}$ \\ 
& $R_3$ & $- \frac{\sin \theta_{13} e^{2i\delta} [ 2 \cos 2\theta_{12} \cos 2\theta_{23} \sin \theta_{13} - \sin 2\theta_{12} \sin 2\theta_{23} (e^{-i\delta} + \sin^2 \theta_{13} e^{i\delta}) ]}{\cos^2 \theta_{13} [ 2 \cos^2 \theta_{12} \cos 2\theta_{23} - \sin 2\theta_{12} \sin 2\theta_{23} \sin \theta_{13} e^{i\delta} ]}$ \\ \hline
\multirow{2}{*}{$\mathbf{D_1^R}$} & $R_2$ & $\frac{\sin \theta_{12} (e^{i\delta} \cos \theta_{12} \sin \theta_{23} \sin \theta_{13} + \sin \theta_{12} \cos \theta_{23})}{\cos \theta_{12} (e^{i\delta} \sin \theta_{12} \sin \theta_{23} \sin \theta_{13} - \cos \theta_{12} \cos \theta_{23})}$ \\ 
& $R_3$ & $-\frac{e^{i\delta} \sin \theta_{13} (e^{i\delta} \cos \theta_{12} \sin \theta_{23} \sin \theta_{13} + \sin \theta_{12} \cos \theta_{23})}{\cos \theta_{12} \sin \theta_{23} \cos^2 \theta_{13}}$ \\ \hline
\multirow{2}{*}{$\mathbf{D_2^R}$} & $R_2$ & $\frac{\sin \theta_{12} (e^{i\delta} \cos \theta_{12} \cos \theta_{23} \sin \theta_{13} - \sin \theta_{12} \sin \theta_{23})}{\cos \theta_{12} (e^{i\delta} \sin \theta_{12} \cos \theta_{23} \sin \theta_{13} + \cos \theta_{12} \sin \theta_{23})}$ \\
& $R_3$ & $\frac{e^{i\delta} \sin \theta_{13} (-e^{i\delta} \cos \theta_{12} \cos \theta_{23} \sin \theta_{13} + \sin \theta_{12} \sin \theta_{23})}{\cos \theta_{12} \cos \theta_{23} \cos^2 \theta_{13}}$ \\ \hline \hline
\end{tabular}
\end{table}

\subsection{Equation for $\cos \delta$}
\label{sec:cub}

In this subsection, we show cubic and quadratic equations whose real solution in terms of $x$ gives $\cos \delta$.

In the $\mathbf{C^R}$ case, the Dirac CP phase can be obtained from the following cubic equation:
\begin{align}
\label{eq:Ceqofdelta}
   &s_{13}^2 \left[ 4 s_{13}^2 \cos^2 2\theta_{12} \cos^2 2\theta_{23} - s_{13} \sin 4\theta_{12} (1+s_{13}^2) x \right. \nonumber \\
   &\left. + \sin^2 2\theta_{12} \sin^2 2\theta_{23} (c_{13}^4 + 4s_{13}^2 x^2) \right] \left[ 2 \left( 2\cos 2\theta_{12} \cos^2 2\theta_{23} - s_{13} \sin 2\theta_{12} \sin 4\theta_{23} x \right) \right] \nonumber \\
   &- \epsilon \left[ 4s_{12}^4 \cos^2 2\theta_{23} + s_{13}^2 \sin^2 2\theta_{23} + 4s_{12}^3 c_{12} s_{13} \sin 4\theta_{23} x \right] \nonumber \\
   &\times \left[ 4\cos^2 2\theta_{23} \left( c_{12}^4 c_{13}^4 - s_{13}^4 \cos^2 2\theta_{12} \right) - s_{13} \sin 4\theta_{23} \left\{ 4c_{13}^4 c_{12}^3 s_{12} - s_{13}^2 \sin 4\theta_{12} \left( 1 + s_{13}^2 \right) \right\} x \right. \nonumber \\
   &\left. - 4s_{13}^4 \sin^2 2\theta_{12} \sin^2 2\theta_{23} x^2 \right] = 0~,
\end{align}
where
\begin{align}
   \epsilon \equiv \frac{\delta m^2}{\Delta m^2 + \delta m^2 / 2}~.
\end{align}

In the $\mathbf{D_1^R}$ case, the Dirac CP phase can be obtained from the following quadratic equation:
\begin{align}
\label{eq:D1eqofdelta}
   &2 c_{23} s_{13}^2 ( 2 c_{12}^2 s_{23}^2 s_{13}^2 + 2 s_{12}^2 c_{23}^2 + \sin 2\theta_{12} \sin 2\theta_{23} s_{13} x )( \cos 2\theta_{12} c_{23} - \sin 2\theta_{12} s_{23} s_{13} x ) \nonumber \\
   &- \epsilon s_{12}^2 ( 2c_{12}^2 s_{23}^2 s_{13}^2 + 2s_{12}^2 c_{23}^2 + \sin 2\theta_{12} \sin 2\theta_{23} s_{13} x ) \nonumber \\
   &\times ( 2c_{12}^2 s_{23}^2 \cos 2\theta_{13} - 2s_{12}^2 c_{23}^2 s_{13}^2 - \sin 2\theta_{12} \sin 2\theta_{23} s_{13}^3 x) = 0~.
\end{align}
This equation can be solved easily and one of the solutions is certainly greater than one. The other solution can give the Dirac CP phase $\delta$ if, and only if it satisfies $-1 \le x \le 1 $:
\begin{align}
   x = \frac{2 \left\{ \cos 2\theta_{12} c_{23}^2 s_{13}^2 + \epsilon s_{12}^2 (s_{12}^2 c_{23}^2 s_{13}^2 - c_{12}^2 s_{23}^2 \cos 2\theta_{13}) \right\}}{\sin 2\theta_{12} \sin 2\theta_{23} s_{13}^3 (1 - \epsilon s_{12}^2)}~.
\end{align}

In the $\mathbf{D_2^R}$ case, the Dirac CP phase can be obtained from the following quadratic equation:
\begin{align}
   &2 s_{23} s_{13}^2 ( 2 c_{12}^2 c_{23}^2 s_{13}^2 + 2 s_{12}^2 s_{23}^2 - \sin 2\theta_{12} \sin 2\theta_{23} s_{13} \cos \delta )( \cos 2\theta_{12} s_{23} + \sin 2\theta_{12} c_{23} s_{13} \cos \delta ) \nonumber \\
   &- \epsilon s_{12}^2 ( 2c_{12}^2 c_{23}^2 s_{13}^2 + 2s_{12}^2 s_{23}^2 - \sin 2\theta_{12} \sin 2\theta_{23} s_{13} \cos \delta ) \nonumber \\
   &\times ( 2c_{12}^2 c_{23}^2 \cos 2\theta_{13} - 2s_{12}^2 s_{23}^2 s_{13}^2 + \sin 2\theta_{12} \sin 2\theta_{23} s_{13}^3 \cos \delta) = 0~.
\end{align}
This equation can be solved easily and one of the solutions is certainly greater than one. The other solution can give the Dirac CP phase $\delta$ if, and only if it satisfies $-1 \le x \le 1 $:
\begin{align}
   x = -\frac{2 \left\{ \cos 2\theta_{12} s_{23}^2 s_{13}^2 + \epsilon s_{12}^2 (s_{12}^2 s_{23}^2 s_{13}^2 - c_{12}^2 c_{23}^2 \cos 2\theta_{13}) \right\}}{\sin 2\theta_{12} \sin 2\theta_{23} s_{13}^3 (1 - \epsilon s_{12}^2)}~.
\end{align}

\subsection{Neutrino mass $m_1$}
\label{app:m1}

Here, we show the list of the formulae of the neutrino mass $m_1$ in Tab.~\ref{tab:m1}.

\begin{table}[h]
\centering
\caption{
The list of the formulae of the neutrino mass $m_1$.
}
\label{tab:m1}
\vspace{5pt}
\begin{tabular}{|c|c|c|} \hline \hline
structure index & $m_1$  \\ \hline 
$\mathbf{C^R}$ & $\delta m \left[ \frac{4s_{12}^2 \cos^22\theta_{23}+4s_{12}^3c_{12}s_{13}\sin4\theta_{23}\cos \delta+s_{13}^2 \sin^22\theta_{12}\sin^22\theta_{23}}{2(2\cos2\theta_{12}\cos^22\theta_{23}-s_{13}\sin2\theta_{12}\sin4\theta_{23}\cos \delta)} \right]^{\frac{1}{2}}$ \\ \hline
$\mathbf{D_1^R}$ & $\delta m \left[ \frac{s_{12}^2 ( 2c_{12}^2 s_{23}^2 s_{13}^2 + 2s_{12}^2 c _{23}^2 + \sin 2\theta_{12} \sin 2\theta_{23} s_{13} \cos \delta)}{2 c_{23} ( \cos 2\theta_{12} c_{23} - \sin 2\theta_{12} s_{23} s_{13} \cos \delta)} \right]^{\frac{1}{2}}$ \\ \hline
$\mathbf{D_2^R}$ & $\delta m \left[ \frac{s_{12}^2 ( 2c_{12}^2 c_{23}^2 s_{13}^2 + 2s_{12}^2 s _{23}^2 - \sin 2\theta_{12} \sin 2\theta_{23} s_{13} \cos \delta)}{2 s_{23} ( \cos 2\theta_{12} s_{23} + \sin 2\theta_{12} c_{23} s_{13} \cos \delta)} \right]^{\frac{1}{2}}$  \\ \hline \hline 
\end{tabular}
\end{table}

\newpage
{\small 
\bibliographystyle{JHEP}
\bibliography{ref}
}

\end{document}